\documentclass[%
 aps,
 prapplied,
 reprint,
 superscriptaddress,
 longbibliography,
 nofootinbib,
 amsmath,amssymb,
 floatfix]{revtex4-2}

\usepackage{graphicx}
\usepackage{booktabs}
\usepackage{siunitx}
\usepackage{tikz}
\usetikzlibrary{positioning, arrows.meta, fit, backgrounds, calc, decorations.pathreplacing}

\newcommand{\venom}{\textsc{Venom}}
\newcommand{\mkid}{MKID}
\newcommand{\mkids}{MKIDs}
\newcommand{\iq}{IQ}
\newcommand{\ssm}{SSM}
\newcommand{\fwhm}{\mathrm{FWHM}}
\newcommand{\dmodel}{d_\mathrm{model}}
\newcommand{\dstate}{d_\mathrm{state}}
\newcommand{\dinner}{d_\mathrm{inner}}
\newcommand{\dconv}{d_\mathrm{conv}}

\newcommand{\kHz}{\,\text{kHz}}
\DeclareMathOperator{\softplus}{softplus}
\DeclareMathOperator{\softmax}{softmax}
\DeclareMathOperator{\SiLU}{SiLU}
\DeclareMathOperator{\GELU}{GELU}
\DeclareMathOperator{\diag}{diag}

\begin{document}

\title{Selective state space model for photon energy estimation in microwave kinetic inductance detectors}

\author{Benjamin A.\ Mazin}
\email[Contact author: ]{bmazin@ucsb.edu}
\affiliation{Department of Physics, University of California, Santa Barbara, California 93106, USA}


\date{\today}

\begin{abstract}
Microwave Kinetic Inductance Detectors (\mkids{}) are superconducting photon counting detectors that simultaneously measure the energy and arrival time of individual photons and are suitable for large-format arrays. The energy resolving power $R = E/\Delta E$ is the primary figure of merit for spectrophotometric applications. It is limited in practice by the assumptions underlying the Wiener optimal filter used to estimate photon energy: stationary noise, linear response, and energy independent pulse shapes, all of which \mkids{} structurally violate. We present \venom{} (Very Efficient Neural Optimal-filter for \mkids{}), a selective state space model based on the Mamba architecture that replaces the coordinate transform and optimal filter, operating directly on raw in-phase/quadrature timestream data, carrying 3{,}314 trainable parameters, and targeting deployment on the MKIDGen3 and future system-on-chip readout platforms. To deal with the small calibration datasets typical of \mkid{} experiments, we train on synthetic pulses drawn from a streaming principal component analysis that samples photon energy continuously between calibration wavelengths. All numbers are reported on a stratified $15\%$ held-out validation set, with $R$ measured from a kernel density estimate. On an InHf bilayer resonator (ten wavelengths, 254--\SI{1310}{nm}), \venom{} reaches a mean $R$ of 26.4 versus 24.2 for the published per-wavelength optimal filter baseline, a $9\%$ improvement from one model in place of ten per-laser filter templates. The $R$ of 38.6 at 254 nm is the highest ever recorded for an ultraviolet-to-near-infrared \mkid{} suitable for use in a dense array. On a PtSi resonator (five wavelengths, 808--\SI{1310}{nm}), where the published analysis used one shared \SI{920}{nm} template, \venom{} matches the optimal filter, with a mean $R$ of 9.1 versus 8.8 at the three interior wavelengths.
\end{abstract}

\maketitle

\section{Introduction} \label{sec:intro}

Microwave Kinetic Inductance Detectors, or \mkids{} \citep{Day2003}, are superconducting photon counting detectors that exploit the change in surface impedance of a superconductor when Cooper pairs are broken by absorbed photons.
Each detector pixel is a lithographically patterned microwave resonator.
Thousands of resonators can be frequency multiplexed onto a single feedline \citep{Mazin2013}, so \mkids{} scale to large-format arrays more readily than most competing single-photon detector technologies.
When a photon is absorbed, the resulting quasiparticles shift the resonant frequency and broaden the resonance.
This appears as a transient perturbation in the complex transmission $S_{21}$ measured at the probe tone frequency.
The magnitude of this perturbation encodes the photon energy, and the energy resolving power
\begin{equation}\label{eq:R}
  R \;\equiv\; \frac{E}{\Delta E} \;=\; \frac{E}{\fwhm}
\end{equation}
is the primary figure of merit for spectrophotometric applications such as exoplanet characterization and high-contrast imaging \citep{Mazin2009, Mazin2013}.
Throughout this work we report $\fwhm$ directly from a kernel density estimate (KDE) of the predicted energy distribution, following the convention established by \citet{Zobrist2022}.

The standard energy estimation pipeline for \mkids{} proceeds in several stages: characterize the resonator's \iq{} loop, apply a coordinate transform to linearize the curved response, construct a Wiener optimal filter from the noise power spectral density and an average pulse template, and calibrate the filter output to photon energy using laser measurements at known wavelengths with a quadratic fit \citep{Zobrist2019, Steiger2022}.
\citet{Zobrist2021} demonstrated that a two-dimensional quadratic coordinate transform can significantly improve the energy resolution by partially compensating for the nonlinear \iq{} geometry.

Despite these advances, the optimal filter remains ``optimal'' only under assumptions that \mkids{} structurally violate: stationary noise, a linear (energy independent) pulse shape, and isolated single-photon events.
The \mkid{} noise environment includes readout power dependent $1/f$ components and two-level system (TLS) fluctuations whose statistics change during and after a photon pulse \citep{Gao2007}.
Higher energy photons produce larger perturbations that explore more of the resonance circle's curvature, so the effective pulse shape is energy dependent.
Stochastic phonon escape into the substrate introduces pulse-to-pulse shape variations correlated with energy \citep{Zobrist2022}.
\citet{Miller2021} showed that principal component analysis (PCA) of pulse shapes captures variations beyond a single matched filter template, improving energy estimation near detector saturation; in the X-ray microcalorimeter community, \citet{Fowler2016} demonstrated a similar gain from ``tangent filtering'' that accounts for energy dependent shapes.

These limitations motivate a learned approach that can capture the full nonlinear, non-stationary response of an \mkid{} directly.
In this paper we present \venom{} (Very Efficient Neural Optimal-filter for \mkids{}), a selective state space model \citep{Gu2023} that replaces the coordinate transform and matched filter with a single neural network operating on raw \iq{} samples.
Three features distinguish \venom{} from prior learned approaches to detector readout~\citep{Ichinohe2022,Fantini2022,makhrinov2025,Rivasto2026}:

\begin{enumerate}
  \item \textbf{FPGA-minimum architecture.} The model contains 3{,}314 parameters arranged so that its recurrent mode inference maps onto a streaming pipeline compatible with the MKIDGen3 RFSoC \citep{Smith2024}. A dual-mode training formulation (Section~\ref{sec:modes}) lets us train in parallel on a GPU and deploy in recurrent mode without retraining.
  \item \textbf{PCA-based synthetic training data.} Calibration datasets for \mkids{} typically contain only $10^4$--$10^5$ pulses per wavelength, at a handful of discrete laser wavelengths. We amortize these small discrete energy sets by fitting a memory efficient PCA model to the calibration pulses and drawing synthetic pulses at continuous energies using a shape preserving (PCHIP) covariance interpolator (Section~\ref{sec:synthetic}).
  \item \textbf{Two-material validation.} We evaluate the same 3{,}314-parameter model on two \mkid{} datasets with substantially different noise properties, pulse timescales, and wavelength ranges: an InHf bilayer resonator (ten wavelengths, 254--\SI{1310}{nm}; \cite{Zobrist2022}) and a PtSi resonator (five wavelengths, 808--\SI{1310}{nm}; \cite{Zobrist2019}).
\end{enumerate}

Section~\ref{sec:background} reviews the relevant \mkid{} physics, the matched filter formalism, and state space models.
Section~\ref{sec:architecture} describes the \venom{} architecture.
Section~\ref{sec:data} details the datasets and preprocessing.
Section~\ref{sec:synthetic} describes the PCA synthetic pulse generator.
Section~\ref{sec:results} presents the energy resolution results for both detectors.
Section~\ref{sec:fpga} discusses the path to FPGA deployment.
Section~\ref{sec:conclusions} summarizes and outlines future work.

\section{Background} \label{sec:background}

\subsection{MKID Physics and IQ Readout} \label{sec:mkid_physics}

\begin{figure}[t]
  \centering
  \includegraphics[width=\columnwidth]{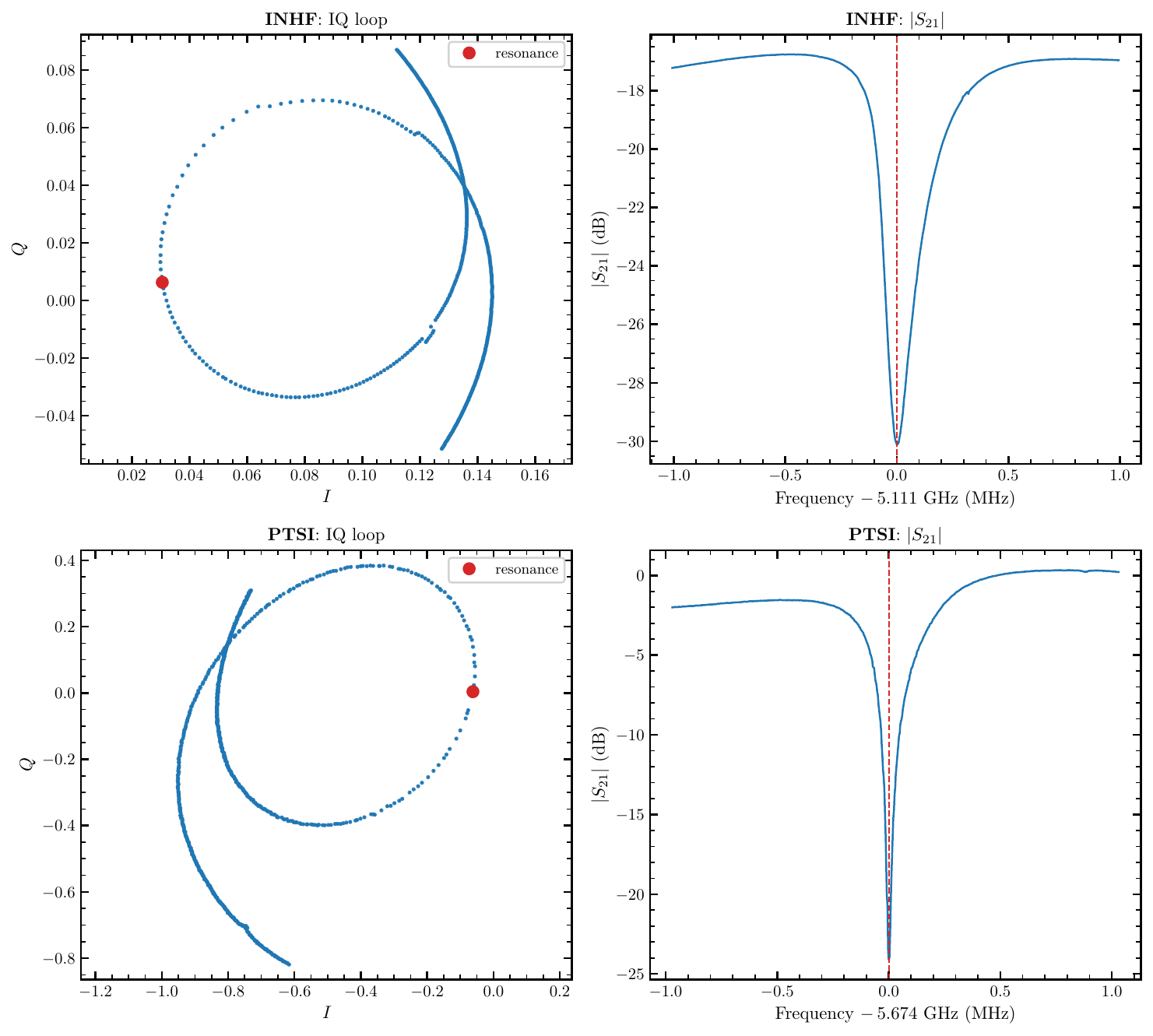}
  \caption{Resonance circle and $|S_{21}|$ transmission for the two test detectors. Top: InHf bilayer resonator near \SI{5.111}{GHz} \citep{Zobrist2022}; bottom: PtSi resonator near \SI{5.674}{GHz} \citep{Zobrist2019}. The red marker indicates the resonance minimum. The asymmetry of each loop arises from impedance mismatch in the feedline coupling.}
  \label{fig:iq_loop}
\end{figure}

Each \mkid{} pixel is a superconducting microresonator probed by a microwave tone at a fixed frequency near the resonance.
The complex transmission coefficient $S_{21}(f)$ traces a circle in the $I$--$Q$ (in-phase/quadrature) plane as the probe frequency is swept through resonance (Figure~\ref{fig:iq_loop}).
At the operating bias frequency, the detector sits at a quiescent point $(I_0, Q_0)$ on this circle.

When a photon with energy $E = hc/\lambda$ is absorbed, Cooper pairs in the superconducting film break into free electrons called quasiparticles, increasing the kinetic inductance and surface resistance.
This shifts the resonant frequency downward and broadens the resonance, causing the observed \iq{} point to move.
The displacement is proportional to the number of quasiparticles created, which in the linear regime is proportional to the photon energy.
The quasiparticles subsequently recombine with a characteristic timescale $\tau_\mathrm{qp}$, which is of order tens of microseconds in PtSi and several hundred microseconds in the InHf bilayer studied here. This produces an exponential ring-down in the \iq{} timestream.

\begin{figure*}[htbp]
  \centering
  \includegraphics[width=\textwidth]{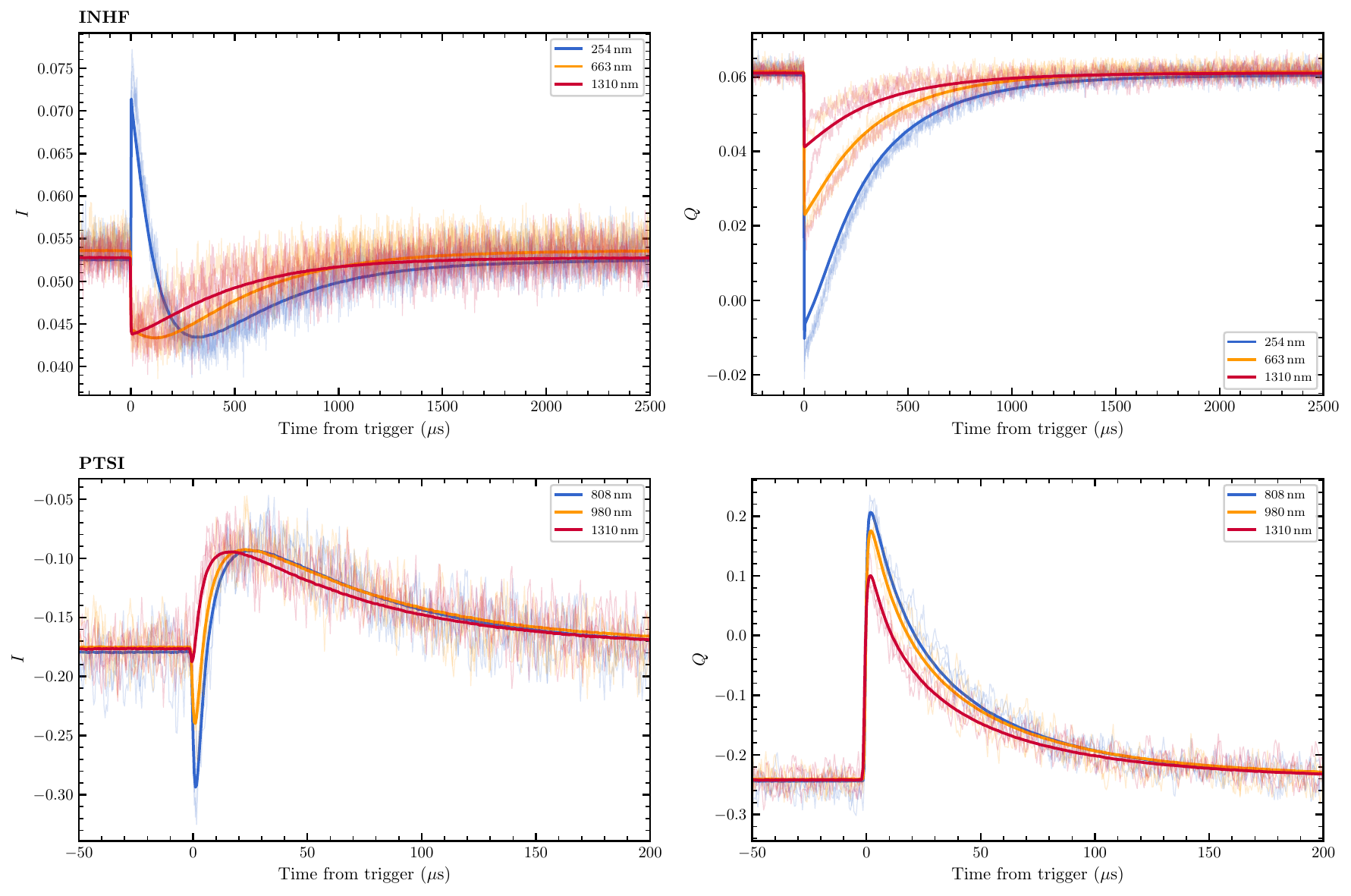}
  \caption{Mean in-phase ($I$) and quadrature ($Q$) pulse traces at three representative wavelengths for each detector, centered on the trigger sample. Thick coloured lines are the cross-pulse mean; thin lines are individual events. Higher energy photons (shorter wavelengths) produce larger pulse amplitudes. Both channels show exponential ring-downs with timescale $\tau_\mathrm{qp}$; the phase and dissipation signals are mixed in the raw $I$, $Q$ basis and require a coordinate transform to separate.}
  \label{fig:traces}
\end{figure*}

Figure~\ref{fig:traces} shows example mean \iq{} traces for representative wavelengths from both datasets.
Higher energy photons produce larger amplitude pulses with longer apparent decay times because the quasiparticle density perturbation scales with photon energy.
The raw $I$ and $Q$ channels both carry mixtures of the phase (angular position on the resonance circle, encoding the frequency shift) and dissipation (radial displacement from the circle center) responses.
Separating these two physical signals requires a coordinate transform that rotates and re-centers the \iq{} loop.
A key motivation for \venom{} is that the model can ignore this decomposition and work directly from the raw \iq{} data, avoiding the need for an explicit transform.

\begin{figure*}[htbp]
  \centering
  \includegraphics[width=\textwidth]{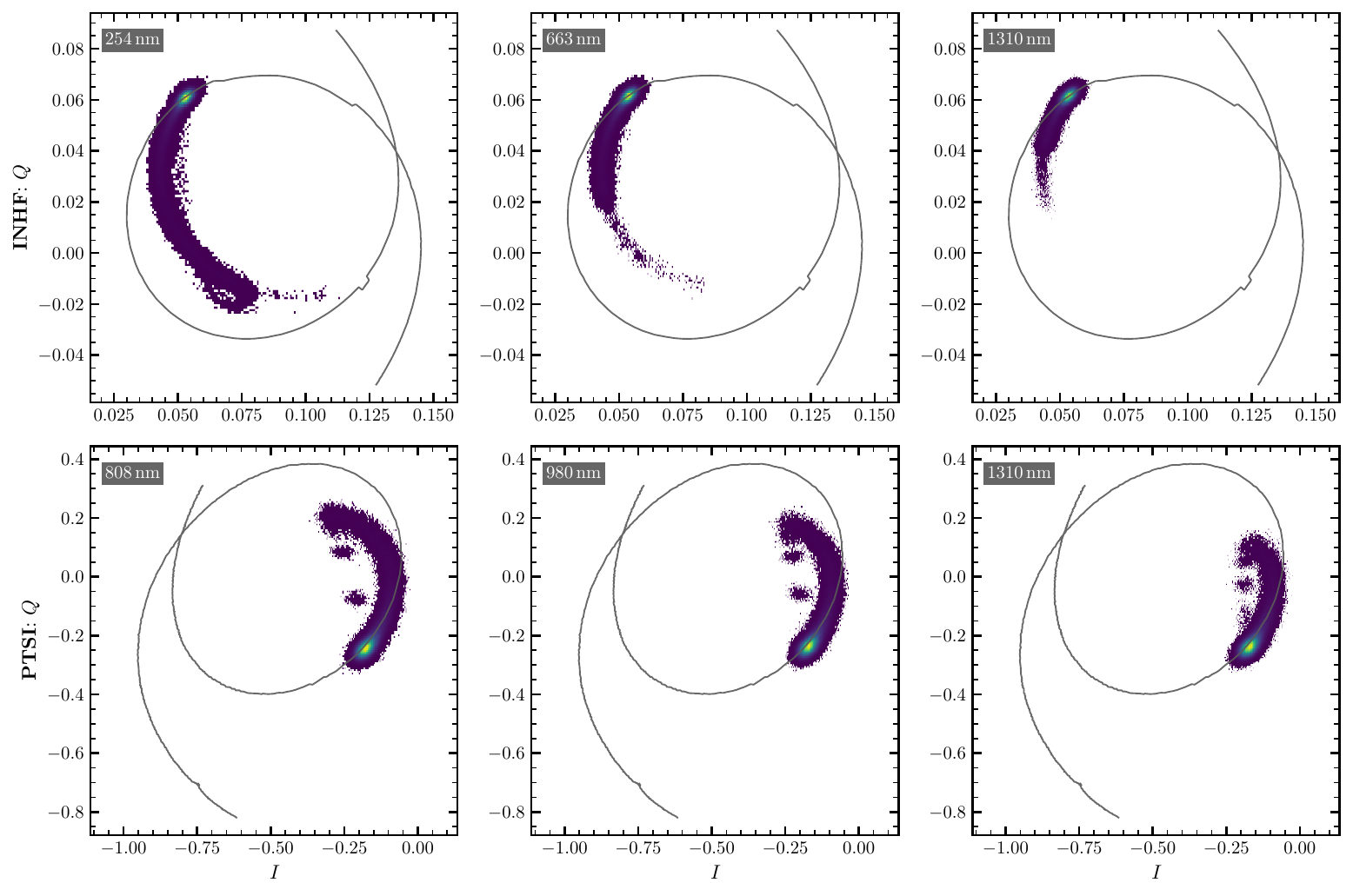}
  \caption{Two-dimensional density of $(I, Q)$ samples from the pulse timestreams at three representative wavelengths, overlaid on the \iq{} resonance loop (grey). Top row: InHf; bottom row: PtSi. Short wavelength pulses excurse further along the loop, exhibiting strong nonlinearity. Long wavelength pulses stay nearer the quiescent point. This energy dependent curvature is not captured by a fixed template optimal filter.}
  \label{fig:trajectories}
\end{figure*}

Crucially, the response is nonlinear in the raw \iq{} coordinates: the photon induced perturbation traces an \emph{arc} that departs from the resonance circle.
Figure~\ref{fig:trajectories} shows the density of pulse samples for representative wavelengths, overlaid on the \iq{} loop.
For the shortest wavelength pulses in both data sets the sample density sweeps a sizable fraction of the loop, well into the regime where the linear approximation breaks down.
This energy dependent nonlinearity is the fundamental reason that a single optimal filter template, trained at one wavelength, cannot be optimal across the full energy range.

\subsection{The Optimal Filter and Its Limitations} \label{sec:matched_filter}

The standard approach to photon energy estimation in \mkids{} is the Wiener filter. This is similar conceptually to the familiar matched filter, but uses the known noise power spectrum to minimize the RMSE between the template and the pulse.

Given a known pulse template $\phi(t)$ and stationary noise with power spectral density $N(f)$, the optimal linear estimator of the pulse amplitude $A$ from a noisy observation $s(t) = A\,\phi(t) + n(t)$ is
\begin{equation}\label{eq:wiener}
  \hat{A} = \frac{\int \tilde{\phi}^*(f)\,\tilde{s}(f)\,/\,N(f)\;\mathrm{d}f}
             {\int |\tilde{\phi}(f)|^2\,/\,N(f)\;\mathrm{d}f},
\end{equation}
where tildes denote Fourier transforms.
This estimator achieves the Cram\'er--Rao bound for Gaussian stationary noise and is therefore ``optimal'' in a precise statistical sense.
For \mkids{}, the filter is typically applied to both the phase (frequency shift) and dissipation quadratures simultaneously as a ``two-component'' filter that combines information from both channels \citep{Zobrist2019}.

The optimality of Equation~\eqref{eq:wiener} relies on four assumptions, each of which \mkids{} violate:
\begin{enumerate}
  \item \textbf{Stationary noise.} \mkid{} noise includes $1/f$ and TLS components whose statistics change when quasiparticles are present.
  \item \textbf{Linear response.} The \iq{} loop geometry introduces curvature (Figure~\ref{fig:trajectories}). \citet{Zobrist2021} partially address this with a quadratic coordinate transform.
  \item \textbf{Energy independent pulse shape.} Higher energy photons explore more of the loop curvature, causing the apparent pulse shape to change with energy.
  \item \textbf{Isolated pulses.} At astrophysical count rates, photon pulses can overlap. The standard matched filter has no mechanism to deconvolve pile-up events, although there are extensions that can address this~\citep{fowler2015,wulf2020}.
\end{enumerate}
Items 2 and 4 admit at least partial remedies within the linear-filter formalism (the quadratic coordinate transform of \citet{Zobrist2021} and the pile-up template extensions of \citet{fowler2015,wulf2020}, respectively). Items 1 and 3 are structural and motivate moving to a learned estimator operating on the raw \iq{} data that can capture the full nonlinear, non-stationary dynamics directly.

\subsection{Scope: One-Resonator Learning, Array Generalisation} \label{sec:scope}

\venom{} is trained \emph{per resonator}: every result quoted in this paper comes from a single 3{,}314-parameter model fit to the calibration pulses of one specific MKID.
The InHf numbers come from the bilayer hafnium-on-indium resonator near \SI{5.111}{GHz} of \citet{Zobrist2022}; the PtSi numbers come from the platinum silicide resonator near \SI{5.674}{GHz} of \citet{Zobrist2019}.
Neither dataset was pooled across multiple sensors.
The 3{,}314 weights therefore describe a learned filter for that resonator's specific noise spectrum, \iq{}-loop geometry, quasiparticle decay constant, and pulse shape distribution, not a universal MKID model.
This is the appropriate apples-to-apples comparison with the published optimal filter baselines, which are themselves built per resonator (a per-wavelength template bank in the InHf case, a single shared template fit on that one resonator in the PtSi case).
The headline claim of this paper is that, on a fixed resonator, a learned sequence model with $\sim$3{,}000 parameters matches or exceeds a hand-built optimal filter pipeline on the same calibration data.

\textbf{Cross-resonator generalization.}
Scaling to thousands-of-resonator MKID arrays, as required for the MKIDGen3 RFSoC platform of \citet{Smith2024}, is left for follow-up work; we sketch two natural paths here.
The first is per-channel training: each resonator gets its own 3{,}314-parameter copy, trained on its own laser calibration. At \SI{4.3}{kB} of W8A32 weights per resonator (Section~\ref{sec:fpga}), this fits comfortably into RFSoC block RAM for arrays of $10^3$--$10^4$ pixels. The second, more interesting path is a shared backbone conditioned on per-channel calibration vectors. This would use $Q_i$, $Q_c$, $f_0$, the \iq{}-loop center and rotation, and possibly the local noise spectrum so that a single trained \venom{} generalizes across many resonators without retraining each one. The PCA synthetic generator of Section~\ref{sec:synthetic} is well suited to this: pulse statistics from many resonators can be fit jointly, with the per-channel vectors entering as additional conditioning inputs to the backbone. We mention this approach as future work in Section~\ref{sec:conclusions}.

\textbf{Cross-time generalisation.}
The matched filter pipeline currently in use for \mkid{} arrays is recalibrated once per cooldown for laboratory measurements. We expect \venom{} to require recalibration on the same cadence and for the same physical reasons: the \iq{}-loop geometry, noise spectrum, and pulse shape that the model has memorized are functions of operating conditions. The PCA synthetic generator can be re-fit on a few minutes of fresh laser data, and the model retrained from scratch in $\lesssim$1 hour per resonator on a single GPU with the recipe of Section~\ref{sec:training}, so the operational cost of refresh is comparable to rebuilding optimal filter templates today. Demonstrating this end-to-end on a multi-cooldown dataset is left for the firmware deployment paper.

\subsection{State Space Models for Sequence Processing} \label{sec:ssm_background}

State space models (\ssm{}s) are a class of sequence-to-sequence models rooted in linear dynamical systems theory.
A continuous-time linear time-invariant \ssm{} maps an input signal $u(t) \in \mathbb{R}$ to an output $y(t) \in \mathbb{R}$ through a latent state $\mathbf{h}(t) \in \mathbb{R}^N$:
\begin{align}
  \frac{\mathrm{d}\mathbf{h}}{\mathrm{d}t} &= \mathbf{A}\,\mathbf{h}(t) + \mathbf{B}\,u(t), \label{eq:ssm_cont_h} \\
  y(t) &= \mathbf{C}\,\mathbf{h}(t) + D\,u(t). \label{eq:ssm_cont_y}
\end{align}
For a diagonal state matrix $\mathbf{A} = \diag(a_1, \ldots, a_N)$ with $a_n < 0$, this system describes $N$ independent exponentially decaying modes, a natural parameterization for physical ring-down processes such as quasiparticle recombination.

Discretizing via the zero-order hold (ZOH) yields the recurrence
\begin{align}
  \mathbf{h}[k] &= \bar{\mathbf{A}}\,\mathbf{h}[k\!-\!1] + \bar{\mathbf{B}}\,u[k], \label{eq:ssm_disc_h} \\
  y[k] &= \mathbf{C}\,\mathbf{h}[k], \label{eq:ssm_disc_y}
\end{align}
with $\bar{\mathbf{A}} = \exp(\mathbf{A}\Delta)$ and $\bar{\mathbf{B}} = \mathbf{A}^{-1}(\bar{\mathbf{A}} - \mathbf{I})\,\mathbf{B}$ (Appendix~\ref{app:zoh}).
This recurrence can be evaluated sample-by-sample (recurrent mode, $\mathcal{O}(1)$ per step) for real-time inference or unrolled as a global convolution (parallel mode) for efficient training on GPUs \citep{Gu2022_S4}.
This dual-mode property is critical: the model can be trained in parallel and deployed in recurrent mode on streaming hardware without retraining.

The Structured State Space for Sequences (S4) model \citep{Gu2022_S4} and its diagonal variant S4D \citep{Gu2022_S4D} demonstrated that \ssm{}s with carefully initialized state matrices can capture long-range dependencies in sequential data.
The Mamba architecture \citep{Gu2023} extended this by making the parameters $\mathbf{B}$, $\mathbf{C}$, and $\Delta$ \emph{input dependent}.
This ``selective'' mechanism allows the model to adapt its filtering dynamically based on the current input, behaving like an infinite impulse response (IIR) filter with time varying, signal adaptive coefficients.
This adaptivity is well suited to \mkid{} data, where the model should behave differently during quiescent noise, during the rising edge of a photon pulse, and during the quasiparticle decay tail.

\section{The VENOM Architecture} \label{sec:architecture}

The next three subsections specify the model's functional form, consisting of a small two-tier neural network whose components have direct physical interpretations. Section~\ref{sec:training} specifies how its weights are learned.
A note on terminology before we begin: every layer described below contains a number of free parameters (matrices, biases, channel-wise scales) that are not designed by hand.
All 3{,}314 of them are jointly fit to training data by minimising a loss function via stochastic gradient descent with the AdamW optimiser \citep{Loshchilov2019}, which adapts each parameter's step size from gradient running statistics; the recipe is fully specified in Section~\ref{sec:training}.
This is the standard ``ML'' paradigm: the architecture below sets up a parameterised function class, and training picks the member of that class that best reproduces the input--output pairs in the calibration data.
Readers unfamiliar with this paradigm can take the parameter counts and matrix shapes below as placeholders to be filled in by training; nothing in the architecture is hand-tuned to a specific resonator.

\subsection{Design Philosophy} \label{sec:design}

\venom{} is designed around three requirements:
(1) operate directly on raw \iq{} samples without hand-crafted coordinate transforms,
(2) learn temporal dynamics on timescales matched to \mkid{} physics, and
(3) be small enough to deploy on FPGA readout electronics at the ADC sample rate with the weights of many channels held simultaneously in block RAM.

\begin{figure}[t]
\centering
\resizebox{\columnwidth}{!}{%
\begin{tikzpicture}[
    >=Stealth,
    block/.style={draw, rounded corners, minimum height=0.7cm, minimum width=1.8cm, align=center, font=\small},
    smallblock/.style={draw, rounded corners, minimum height=0.55cm, minimum width=1.4cm, align=center, font=\scriptsize},
    arr/.style={->, thick},
    label/.style={font=\scriptsize\itshape, text=gray},
    tier/.style={draw, dashed, rounded corners=4pt, inner sep=6pt},
  ]
  \node[block, fill=blue!8] (input) {$I[k],\; Q[k]$};
  \node[label, below=0pt of input] {$\mathbb{R}^2$ at ADC rate};

  \node[block, fill=orange!12, above=0.6cm of input] (proj) {Linear $(2 \!\to\! 16)$};
  \draw[arr] (input) -- (proj);

  \node[block, fill=red!10, above=0.5cm of proj, minimum width=2.5cm] (mb1) {MambaBlock};
  \draw[arr] (proj) -- (mb1);

  \node[smallblock, fill=gray!8, above=0.35cm of mb1] (norm) {LayerNorm};
  \draw[arr] (mb1) -- (norm);

  \begin{scope}[on background layer]
    \node[tier, fill=red!3, fit=(proj)(mb1)(norm), label={[font=\small\bfseries, text=red!70!black]left:Tier 1}] (t1) {};
  \end{scope}

  \node[label, right=0.6cm of norm] {$\mathbf{f}[k] \in \mathbb{R}^{16}$};

  \node[block, fill=green!12, above=0.7cm of norm] (attn) {Attention Pool};
  \draw[arr] (norm) -- (attn);

  \node[block, fill=green!12, above=0.4cm of attn] (mlp) {MLP $(16\!\to\!32\!\to\!32\!\to\!2)$};
  \draw[arr] (attn) -- (mlp);

  \node[block, fill=blue!8, above=0.5cm of mlp] (output) {$(\hat{\mu},\;\log\hat\sigma^2)$};
  \draw[arr] (mlp) -- (output);

  \begin{scope}[on background layer]
    \node[tier, fill=green!3, fit=(attn)(mlp)(output), label={[font=\small\bfseries, text=green!50!black]left:Tier 2}] (t2) {};
  \end{scope}

  \node[label, right=0.6cm of t1.east, text width=1.8cm, align=left] {ADC rate};
  \node[label, right=0.6cm of t2.east, text width=1.8cm, align=left] {Photon rate\\($\sim$\kHz{})};

  \node[anchor=north east, font=\scriptsize, text width=3.2cm, align=left] at ([xshift=-0.4cm]t1.north west) {
    \textbf{Each MambaBlock:}\\
    LayerNorm\\
    $\to$ $\mathrm{in\_proj}$ ($16\!\to\!32$),\\
    \ \ split $\to$ $\mathbf{x} \,|\, \text{gate}$ ($16\,|\,16$)\\
    $\to$ Conv1D($k\!=\!4$) + SiLU on $\mathbf{x}$\\
    $\to$ Selective SSM($N\!=\!8$)\\
    $\to$ $\odot\;\SiLU(\text{gate})$\\
    $\to$ $\mathrm{out\_proj}$ ($16\!\to\!16$) + residual
  };

\end{tikzpicture}
}%
\caption{The \venom{} architecture. Tier~1 (SSM backbone) processes raw IQ samples at the ADC rate via a single Mamba block, producing a 16-dimensional feature stream. Tier~2 (energy head) pools features over the photon event window using learned attention weights and regresses a Gaussian mean and log variance via an MLP. At inference, Tier~1 runs in recurrent mode (one sample at a time); Tier~2 runs once per triggered photon.}
\label{fig:architecture}
\end{figure}
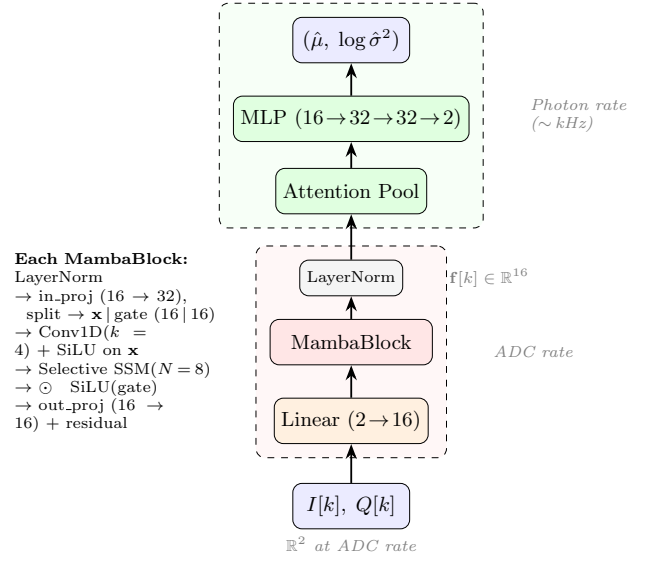

These requirements motivate a two-tier architecture (Figure~\ref{fig:architecture}).
\textbf{Tier~1} (the \ssm{} backbone) processes raw $I[k], Q[k]$ samples in recurrent mode, producing a latent feature stream at the downsampled FGPA output rate.
This tier replaces both the coordinate transform and the matched filter, implicitly learning a nonlinear, noise adaptive filtering.
\textbf{Tier~2} (the energy estimating model, referred to as the energy head) takes a window of Tier~1 features around a triggered photon event and regresses to photon energy.
It runs at the photon rate ($\sim$\kHz{}), amortizing its computational cost over many ADC samples.

\subsection{Tier 1: SSM Backbone} \label{sec:backbone}

The backbone takes as input the two-channel \iq{} timestream $\mathbf{u}[k] = (I[k],\; Q[k])^T \in \mathbb{R}^2$ at each sample $k$.
A linear input projection maps $\mathbf{u}[k]$ to a $\dmodel = 16$-dimensional feature vector:
\begin{equation}
  \mathbf{x}^{(0)}[k] = \mathbf{W}_\mathrm{in}\,\mathbf{u}[k] + \mathbf{b}_\mathrm{in}, \quad \mathbf{W}_\mathrm{in} \in \mathbb{R}^{\dmodel \times 2}.
\end{equation}
This embedding does not, by itself, add information beyond the two facts $(I[k], Q[k])$; it just reshapes them to match the Mamba block's internal width, where the actual feature extraction happens. The block has 16 channels because that width was selected by the architecture sweep (Section~\ref{sec:arch_sweep}) as the smallest $\dmodel$ that retains full energy resolution while fitting the FPGA budget.
This feature stream then passes through $L = 1$ Mamba block.
A single block backbone is sufficient here because the selective \ssm{} state captures the relevant ring-down dynamics directly. Adding a second block changes mean $R$ by $-0.1$ on PtSi and $+0.8$ on InHf (Table~\ref{tab:arch_sweep}) at $+47\%$ parameters, a marginal trade once the FPGA cost of an extra block is factored in.
The block applies layer normalization (a per-sample rescaling that fixes the channel-wise mean and variance, stabilising the gradient through the block), expands the feature dimension by a unit factor ($\dinner = \dmodel$), splits the expanded representation into two branches, and applies a gated selective \ssm{}:

\begin{enumerate}
  \item \textbf{Split.} A linear projection maps the normalized input to $2\dinner$ dimensions, which are split into an \ssm{} branch $\mathbf{x}_\mathrm{ssm}$ and a gate branch $\mathbf{z}$, each $\in \mathbb{R}^{\dinner}$.
  \item \textbf{Local convolution.} A causal 1D convolution with kernel size $\dconv = 4$ is applied depthwise to $\mathbf{x}_\mathrm{ssm}$, followed by a SiLU activation (a smooth sigmoid weighted nonlinearity, $\SiLU(x) = x\,\sigma(x)$, that approximates ReLU but is differentiable everywhere). This 4-sample kernel covers the fast rising edge of photon pulses at the effective sample rate.
  \item \textbf{Selective \ssm{}.} The convolved features are processed by a diagonal selective \ssm{} with state dimension $\dstate = 8$. Selectivity adapts the filter coefficients sample-by-sample across pulse regimes (quiescent baseline, rising edge, decay tail), rather than holding them fixed as in a classical IIR filter, so the same block can integrate the baseline, sharpen the rising edge, and average over the decay tail. ``Input dependent'' is the central distinction from a classical fixed-coefficient IIR filter: the selection vectors $\mathbf{B}[k]$, $\mathbf{C}[k]$ and the discretisation timestep $\boldsymbol{\Delta}[k]$ are not learned constants but are computed at each sample as learned linear projections of the current feature vector $\mathbf{x}[k]$,
  \begin{align}
    \mathbf{B}[k] &= \mathbf{W}_B\,\mathbf{x}[k], \quad \mathbf{C}[k] = \mathbf{W}_C\,\mathbf{x}[k], \\
    \boldsymbol{\Delta}[k] &= \softplus(\mathbf{W}_\Delta\,\mathbf{x}[k] + \mathbf{b}_\Delta),
  \end{align}
  so the filter coefficients adapt sample by sample to the local pulse phase (quiescent baseline, rising edge, decay tail).
  What is learned here are the projection matrices $\mathbf{W}_B, \mathbf{W}_C \in \mathbb{R}^{\dstate \times \dinner}$ and $\mathbf{W}_\Delta \in \mathbb{R}^{\dinner \times \dinner}$ together with the bias $\mathbf{b}_\Delta$. Their outputs $\mathbf{B}[k]$ and $\mathbf{C}[k]$ are vectors, one pair per sample, and are matrices only in the degenerate $\mathbb{R}^{N \times 1}$ sense of the single-input system of Equations~\eqref{eq:ssm_cont_h}--\eqref{eq:ssm_cont_y}.
  The shapes carried through the block are $\mathbf{x}[k] \in \mathbb{R}^{\dinner}$, $\mathbf{B}[k], \mathbf{C}[k] \in \mathbb{R}^{\dstate}$, $\boldsymbol{\Delta}[k] \in \mathbb{R}^{\dinner}$, $\mathbf{A} \in \mathbb{R}^{\dinner \times \dstate}$, $\mathbf{D} \in \mathbb{R}^{\dinner}$, and a hidden state $\mathbf{h}[k] \in \mathbb{R}^{\dinner \times \dstate}$ of $16 \times 8 = 128$ floats, which is the per-layer state carried between samples in recurrent mode.
  The state matrix $\mathbf{A}$ holds one decay rate $a_{i,n}$ for each pairing of channel $i$ with mode $n$ and is shared across timesteps, with $a_{i,n} < 0$ enforced by the parameterization $a_{i,n} = -\exp(\theta_{i,n})$.
  Every channel carries its own $\dstate$-dimensional state, and with the discretised decay written as $\bar{A}_{i,n}[k] \equiv e^{a_{i,n} \Delta_i[k]}$ the update is elementwise over the mode index,
  \begin{equation}\label{eq:state_update}
    \boxed{
    \begin{aligned}
      h_{i,n}[k] &= \bar{A}_{i,n}[k]\,h_{i,n}[k\!-\!1] \\
                 &\quad + \frac{\bar{A}_{i,n}[k] - 1}{a_{i,n}}\,B_n[k]\,x_i[k],
    \end{aligned}}
  \end{equation}
  for $i = 1, \ldots, \dinner$ and $n = 1, \ldots, \dstate$, with channel output
  \begin{equation}\label{eq:ssm_out}
    y_i[k] = \sum_{n=1}^{\dstate} C_n[k]\,h_{i,n}[k] + D_i\,x_i[k],
  \end{equation}
  equivalently $\mathbf{C}[k]^T \mathbf{h}_i[k] + D_i\,x_i[k]$ in terms of the channel state vector $\mathbf{h}_i[k]$.
  Note that $\mathbf{B}[k]$ and $\mathbf{C}[k]$ carry no channel index. The same pair of selection vectors is shared by all $\dinner$ channels at a given sample, whereas $\boldsymbol{\Delta}[k]$, $\mathbf{D}$, and the rows of $\mathbf{A}$ are per channel.
  \item \textbf{Gating and residual.} The \ssm{} output is element-wise multiplied by $\SiLU(\mathbf{z})$, projected back to $\dmodel$ dimensions, and added to the block input via a residual connection.
\end{enumerate}

A final layer normalization produces the backbone feature stream $\mathbf{f}[k] \in \mathbb{R}^{\dmodel}$.

\textbf{Initialization.}
The entries of $\mathbf{A}$ are initialized following the S4D-Real scheme \citep{Gu2022_S4D}: $a_n = -\exp(\mathrm{linspace}(\ln 1, \ln \dstate, \dstate))$ for $n = 1, \ldots, \dstate$, broadcast identically across the $\dinner$ channels. Training therefore starts from a single shared mode spectrum and lets the channels specialize away from it.
The bias of the $\boldsymbol{\Delta}$ projection is initialized uniformly in $[\ln(0.001), \ln(0.1)]$, yielding initial discretization timesteps that bracket the range of physical ring-down timescales observed across both datasets.

\subsection{Tier 2: Energy Head} \label{sec:head}

The energy head estimates the photon energy from a window of backbone features $\{\mathbf{f}[k]\}_{k=1}^{T}$ surrounding a triggered event.
Rather than averaging all timesteps equally, it uses a learned attention weighted temporal pool: a softmax normalised weighted sum of the per-sample features, where the weights themselves are computed from those same features through a learned linear scoring vector $\mathbf{w}$,
\begin{align}
  \alpha[k] &= \softmax_k\!\bigl(\mathbf{w}^T \mathbf{f}[k]\bigr), \label{eq:attn_weight}\\
  \mathbf{z} &= \sum_{k=1}^{T} \alpha[k]\,\mathbf{f}[k]. \label{eq:attn_pool}
\end{align}
The softmax operator $\softmax_k(s_k) = e^{s_k} / \sum_{k'} e^{s_{k'}}$ converts the raw scores into non-negative weights that sum to one over the window.
This pooling lets the model concentrate on informative timesteps, typically the pulse peak and early decay, while down-weighting baseline and late time noise.

The pooled vector $\mathbf{z} \in \mathbb{R}^{\dmodel}$ is passed through a two-hidden-layer multilayer perceptron (MLP -- a stack of dense linear layers separated by element-wise nonlinearities) with GELU activations (a smooth ReLU-like function, $\GELU(x) = x\,\Phi(x)$ where $\Phi$ is the standard normal CDF) and hidden dimension $32$, producing two outputs: an energy estimate $\hat{\mu}$ and a log variance $\log \hat{\sigma}^2$.
Predicting $\hat\sigma^2$ alongside $\hat\mu$ lets the loss down-weight pulses the model cannot fit (pile-up tails, multi-photon coincidences, baseline glitches), so a small population of unfit pulses does not bias the bulk regression on $\hat\mu$.
We train end-to-end with the Gaussian negative log-likelihood (NLL) loss
\begin{equation}\label{eq:loss}
  \mathcal{L} = \frac{1}{N_\mathrm{batch}} \sum_{j=1}^{N_\mathrm{batch}}
    \frac{1}{2}\!\left[\log \hat{\sigma}_j^2 + \frac{(\hat{\mu}_j - E_j)^2}{\hat{\sigma}_j^2}\right],
\end{equation}
and read the energy estimate directly from $\hat{\mu}$ at inference. To prevent the well-known variance collapse pathology of Gaussian NLL on easy samples \citep{Seitzer2022}, we clip $\log\hat\sigma^2$ to $[-6, 4]$ during the loss computation.
Table~\ref{tab:loss_ablation} reports a four-way loss ablation $\{$MSE, Huber, Gaussian NLL, $\beta$-NLL$\}$ on PtSi with three seeds each, holding all other hyperparameters at the published winner. The four losses reach the same resolving power within 0.3 in $\bar R$ ($10.11 \pm 0.15$ for MSE, $10.24 \pm 0.15$ for Huber, $10.31 \pm 0.04$ for Gaussian NLL and $10.39 \pm 0.08$ for $\beta$-NLL), and all four are stable across seeds. The loss therefore does not set the resolution of $\hat\mu$ on this dataset, and we keep Gaussian NLL because it also trains the variance head at no cost in $R$. 

\paragraph{Variance head calibration.}
We do not use $\hat\sigma^2$ as a downstream uncertainty in the published recipe, but the head is trained on it, so we audit whether it is calibrated.
On the held-out validation pulses the variance prediction is set by the lower clip floor. $\log\hat\sigma^2$ saturates at $-6$ on $83\%$ of InHf pulses and $34\%$ of PtSi pulses, so on those pulses the predicted $\hat\sigma$ stays constant at \SI{50}{meV}. Across the full validation set the reduced $\chi^2$ of the standardized residuals $z_j = (E_j - \hat\mu_j)/\hat\sigma_j$ is $0.50$ on InHf (over-dispersed, floor too large) and $0.95$ on PtSi, and the probability-integral-transform histogram is dome-shaped on InHf and close to uniform on PtSi. On InHf the head therefore behaves close to homoscedastic, and the Gaussian NLL of Equation~\ref{eq:loss} acts as a robust regression loss with a small admixture of heteroscedasticity concentrated at the highest energy InHf wavelengths (the head drives $\log\hat\sigma^2$ above the clip floor on $\sim$17\% of InHf pulses overall, predominantly at the 254 and \SI{313}{nm} lines). On PtSi it predicts 62--\SI{66}{meV} at 808--\SI{980}{nm} and sits at the floor at \SI{1310}{nm} and on most \SI{1120}{nm} pulses. We quote only $\hat\mu$ throughout the rest of the paper.

\paragraph{Trading resolving power for a calibrated $\hat\sigma$.}
Every number quoted elsewhere in this paper comes from the published recipe, with the clip at $[-6, 4]$ and only $\hat\mu$ read out.
This paragraph asks what it would cost to read out $\hat\sigma$ as well.
Lifting the floor to $\log\hat\sigma^2 = -14$ ($\hat\sigma_{\min} \approx \SI{0.9}{meV}$) removes the saturation completely: the pinned fraction falls from $34\%$ to zero on PtSi and from $83\%$ to zero on InHf, and the per-wavelength median $\hat\sigma$ then spreads over 39--\SI{64}{meV} across the five PtSi lines and 30--\SI{66}{meV} across the ten InHf lines.
This costs no resolution on either detector: mean $R$ moves from 10.3 to 10.5 on PtSi and from 26.2 to 26.5 on InHf (seed 42), within the scatter between seeds.

Table~\ref{tab:clip_sweep} separates the floor from the loss.
Holding $\beta$ fixed and sweeping the floor leaves $\bar R$ unchanged, because below $-6.5$ the floor never binds: the head does not predict $\hat\sigma$ below \SI{18}{meV} on PtSi.
Holding the floor at $-14$ and sweeping $\beta$ over $\{0, 0.1, 0.5, 1.0\}$ moves $\bar R$ by $0.8$ in total, with $\beta = 1$ lowest.
Here $\beta = 0$ is ordinary Gaussian NLL, and $\beta = 1$ gives MSE-equivalent gradients on $\hat\mu$ while still training $\hat\sigma$.
We keep $\beta = 0.5$ by convention rather than because it earns the resolution.

\begin{table}[t]
  \centering
  \caption{PtSi mean KDE resolving power under the two knobs of the variance head. Top: floor sweep at fixed $\beta = 0.5$. Bottom: $\beta$ sweep at fixed floor $-14$. All entries use seed 42 and the same $15\%$ stratified hold-out used throughout. The floor $-6$ entry is the seed 42 run of the $\beta$-NLL row of Table~\ref{tab:loss_ablation}, and the last row is the seed 42 production checkpoint.}
  \label{tab:clip_sweep}
  \begin{tabular}{lcc}
    \toprule
    Configuration & $\hat\sigma_{\min}$ (meV) & PtSi $\bar R$ \\
    \midrule
    \multicolumn{3}{l}{\emph{Floor sweep, $\beta = 0.5$}} \\
    \quad floor $=-6$    & 49.8 & 10.47 \\
    \quad floor $=-6.5$  & 38.8 & 10.45 \\
    \quad floor $=-8$    & 18.3 & 10.47 \\
    \quad floor $=-10$   & 6.7  & 10.47 \\
    \quad floor $=-14$   & 0.9  & 10.47 \\
    \midrule
    \multicolumn{3}{l}{\emph{$\beta$ sweep, floor $=-14$}} \\
    \quad $\beta = 0$ (Gaussian NLL) & 0.9 & 10.36 \\
    \quad $\beta = 0.1$              & 0.9 & 10.36 \\
    \quad $\beta = 0.5$              & 0.9 & 10.47 \\
    \quad $\beta = 1.0$              & 0.9 & 9.67 \\
    \midrule
    Published recipe (Gaussian NLL,\\ floor $=-6$) & 49.8 & 10.29 \\
    \bottomrule
  \end{tabular}
\end{table}

The resulting $\hat\sigma$ is directly usable on both detectors.
Over the whole validation set the standardized residuals $z_j = (E_j - \hat\mu_j)/\hat\sigma_j$ give $\chi^2_\nu = 1.01$ on InHf and $1.04$ on PtSi, with $1$/$2$/$3\sigma$ coverage of $0.697$/$0.952$/$0.994$ and $0.707$/$0.949$/$0.992$ against the Gaussian $0.683$/$0.954$/$0.997$.
No trimming and no rescaling are needed, and applying either degrades the agreement.


For an application that wants a per-pulse uncertainty, then, the recipe is to train the lifted-clip variant and treat $\hat\sigma$ at inference as the $1\sigma$ uncertainty, flagging $|E - \hat\mu| > 3\,\hat\sigma$ as a candidate outlier.
This adds no inference cost and no measurable loss of resolution. We keep the published recipe only so that every other number in the paper comes from one configuration.


\begin{table*}[t]
  \centering
  \small
  \setlength{\tabcolsep}{4pt}
  \renewcommand{\arraystretch}{1.08}
  \caption{Loss-function ablation on PtSi, holding the published-winner architecture, training recipe (25 epochs, 120{,}000 PCA-synthetic pulses/epoch, AdamW), and the same 15\% stratified validation hold-out fixed. Each row reports the per-wavelength KDE $R$ averaged over three seeds; $\bar{R}$ is the mean across the five wavelengths, $\pm$ one standard deviation across seeds. $\hat\mu$ is read out at inference for all losses; the variance head is exercised by Gaussian NLL and $\beta$-NLL only. The Huber transition $\delta = 0.05$ eV is approximately one validation-RMS residual on the winner; we use $\beta = 0.5$ for $\beta$-NLL, the Seitzer recommendation. The optimal filter row gives the resolving powers of the published analysis \citep{Zobrist2019}. $^\ddagger$From our reanalysis of the traces with the published recipe, which reproduces the published values within 2\%.}
  \label{tab:loss_ablation}
  \begin{tabular}{lccccccc}
    \toprule
    Loss & Seeds & 808\,nm & 920\,nm & 980\,nm & 1120\,nm & 1310\,nm & $\bar{R}$ \\
    \midrule
    Optimal Filter & --- & 8.5 & 8.8$^\ddagger$ & 9.0 & 8.8$^\ddagger$ & 8.9 & 8.8 \\
    \midrule
    MSE & 3 & 11.9 & 8.9 & 8.6 & 9.7 & 11.4 & $10.11 \pm 0.15$ \\
    Huber ($\delta{=}0.05$) & 3 & 11.9 & 9.5 & 8.3 & 9.8 & 11.8 & $10.24 \pm 0.15$ \\
    Gaussian NLL & 3 & 12.0 & 9.3 & 8.3 & 9.7 & 12.1 & $10.31 \pm 0.04$ \\
    $\beta$-NLL ($\beta{=}0.5$) & 3 & 11.8 & 9.9 & 8.4 & 9.6 & 12.2 & $10.39 \pm 0.08$ \\
    \bottomrule
  \end{tabular}
\end{table*}

\subsection{Training and Inference Modes} \label{sec:modes}

The \ssm{} recurrence of Equation~\eqref{eq:state_update} can equivalently be evaluated as a parallel cumulative sum scan during training and as the sample-by-sample recurrence during real-time inference. Both paths compute the same function in exact arithmetic but accumulate floating-point error differently. To keep the parallel scan numerically aligned with the recurrent state update, we segment it into chunks of 64 samples with the hidden state carried between segments and use a Hillis--Steele parallel-prefix recurrence inside each segment, so the cumulative product $\prod_t \bar{\mathbf{A}}_t$ is never reformulated as $\exp(\sum_t \log \bar{\mathbf{A}}_t)$ with clipping (an earlier clip-and-exponentiate formulation introduced biases of up to \SI{362}{meV} on InHf pulses whose $\log\bar A_t$ fell outside the clip range; the prefix-sum scan eliminates this bias).
Empirically, the two paths agree to within \SI{2}{\micro eV} on every validation pulse of both datasets (max disagreement \SI{1.9}{\micro eV} on InHf, \SI{0.6}{\micro eV} on PtSi --- $5\times10^{-5}\,\sigma_\|$ and $1\times10^{-5}\,\sigma_\|$ respectively), and the recurrent mode per-wavelength KDE $R$ matches the parallel number to within $|\Delta R| \le 0.03$ on all fifteen wavelengths; the full disagreement distribution is in Appendix~\ref{app:parrec}.
This is a training mode equivalence check, not a deployment guarantee: the dominant fixed-point-hardware risk is quantization of weights and activations to integer arithmetic, addressed in Section~\ref{sec:fpga} (Table~\ref{tab:ptq}), with full activation quantization and hardware emulation deferred to a planned firmware paper.

Table~\ref{tab:hyperparams} summarizes the model hyperparameters and configuration.

\begin{table}[t]
  \centering
  \caption{Model configuration and training hyperparameters. Both datasets use the same recipe, differing only in the preprocessing dependencies of Table~\ref{tab:dataset}.}
  \label{tab:hyperparams}
  \begin{tabular}{ll}
    \toprule
    Parameter & Value \\
    \midrule
    \multicolumn{2}{l}{\textit{Architecture}} \\
    Model dimension ($\dmodel$) & 16 \\
    State dimension ($\dstate$) & 8 \\
    Number of layers ($L$) & 1 \\
    Conv kernel size ($\dconv$) & 4 \\
    Expansion factor & 1 \\
    Head hidden dimension & 32 \\
    Head output dimension & 2 $(\hat\mu,\, \log\hat\sigma^2)$ \\
    Total parameters & 3{,}314 \\
    \midrule
    \multicolumn{2}{l}{\textit{Training}} \\
    Optimizer & AdamW \\
    Peak learning rate & $10^{-3}$ \\
    Weight decay & $10^{-4}$ \\
    LR schedule & Cosine anneal + warmup \\
    Warmup epochs & 5 \\
    Total epochs & 25 (PtSi), 100 (InHf) \\
    Batch size & 256 \\
    Synthetic pulses / epoch & 120{,}000 \\
    Loss function & Gaussian NLL (Equation~\ref{eq:loss}) \\
    \bottomrule
  \end{tabular}
\end{table}

\section{Data and Preprocessing} \label{sec:data}

\subsection{Datasets} \label{sec:dataset}

We evaluate \venom{} on two laser calibration datasets taken with a traveling-wave parametric amplifier (TWPA) readout:

\textbf{InHf bilayer \mkid{}} \citep{Zobrist2022}.
A hafnium-on-indium bilayer resonator operated at an ADC sample rate of \SI{0.8}{MHz}.
Ten laser wavelengths from 254 to \SI{1310}{nm} cover the UV through NIR (photon energies 0.95--\SI{4.88}{eV}).
Quasiparticle recombination is slow ($\tau_\mathrm{qp} \sim \SI{355}{\micro s}$).

\textbf{PtSi \mkid{}} \citep{Zobrist2019}.
A platinum silicide resonator operated at \SI{2.0}{MHz} sample rate.
Five laser wavelengths span 808 to \SI{1310}{nm} (0.95--\SI{1.53}{eV}), the same NIR region that dominates the InHf lower energy coverage.
The PtSi quasiparticle decay is faster ($\tau_\mathrm{qp} \sim \SI{40}{\micro s}$).

The ground truth photon energy for each event is computed from the known laser wavelength via $E = hc/\lambda$.
A pre-computed quality mask \citep{Zobrist2019} rejects events contaminated by pile-up, cosmic rays, or triggering artifacts.
Table~\ref{tab:dataset} summarizes the per-dataset preprocessing; both use an identical recipe parameterized only by the raw sample rate.

\begin{table}[t]
  \centering
  \caption{Dataset summary and shared preprocessing recipe. Both datasets pass through the same pipeline; only the per-dataset columns differ. Shared value rows are marked ``$=$''.}
  \label{tab:dataset}
  \small
  \setlength{\tabcolsep}{4pt}
  \renewcommand{\arraystretch}{1.08}
  \begin{tabular}{lcc}
    \toprule
      & InHf & PtSi \\
    \midrule
    Source & \citeauthor{Zobrist2022} 2022 & \citeauthor{Zobrist2019} 2019\\
    Raw ADC rate & 0.8\,MHz & 2.0\,MHz \\
    Wavelengths & 10 (254--1310\,nm) & 5 (808--1310\,nm) \\
    Energy (eV) & 0.95--4.88 & 0.95--1.53 \\
    $\tau_\mathrm{qp}$ (eff.) & $\sim$355\,$\mu$s & $\sim$40\,$\mu$s \\
    Eff.\ rate & 200\,kHz & 500\,kHz \\
    Pulses after cut & 63{,}568 & 37{,}000 \\
    \midrule
    Downsample        & 4 & $=$ \\
    Window            & $5\tau_\mathrm{qp}$ & $=$ \\
    Peak-$\sigma$ MAD & 2.0 & $=$ \\
    Val split         & 15\% (stratified) & $=$ \\
    \bottomrule
  \end{tabular}
\end{table}

\subsection{Preprocessing} \label{sec:preprocessing}

The preprocessing pipeline is a single function parameterized by the raw sample rate, applied identically to both datasets. The training/validation split of the last step is drawn first, every quantity the pipeline fits comes from the training pulses alone, and no step that transforms a pulse depends on that pulse's wavelength label:

\begin{enumerate}
  \item \textbf{Quality masking.} Events flagged by the quality mask from the original Zobrist analysis are removed.
  \item \textbf{Peak height outlier rejection.} A median-absolute-deviation (MAD) cut at $|\mathrm{PH} - \tilde{\mathrm{PH}}| < 2.0\,\sigma_\mathrm{MAD}$ removes IR-contamination pulses whose two-component optimal filter peak heights lie far below the primary cluster. Each line's median and MAD come from its training pulses, and the cut applies to both halves. This cut is essential at the shortest InHf wavelengths (254, 313, \SI{406}{nm}), where an IR contamination peak can otherwise impact the quoted resolving power by polluting the Gaussian baseline fit.
  \item \textbf{Baseline subtraction.} Each laser line was recorded as a separate run, and the quiescent point of the resonator drifts slightly between runs. Left in the data, that offset would tell the model which line a pulse came from. We therefore subtract from every trace its own baseline, the $3\sigma$ clipped mean of 500 native samples ending 200 samples before the trigger, and move all pulses to one common quiescent point, the clipped mean of the training baselines. One trigger position, estimated from training pulses of all lines, places this window for every pulse.
  \item \textbf{Sub-sample pulse alignment.} One alignment template serves every line: the mean pulse envelope $|z - z_\mathrm{bl}|$ of the training pulses, where $z = I + iQ$ and $z_\mathrm{bl}$ is the pulse's own baseline. Every trace's envelope is cross-correlated against it via FFT within $\pm 5$ native samples; the integer lag peak is refined to sub-sample precision by fitting a parabola to the three samples around the correlation maximum, and the trace is then resampled by the resulting fractional shift using a linear (order-1) interpolating shift to avoid Gibbs ringing. The template is refined by aligning the training pulses to it twice. RMS shifts are 0.15--0.64 native samples per line.
  \item \textbf{Downsampling.} Both datasets are decimated by a factor of 4 using an anti-aliasing FIR filter. 
  \item \textbf{Pulse windowing.} Each trace is cropped to a window of length $n_\tau \tau_\mathrm{qp} = 5\tau_\mathrm{qp}$ past the trigger plus a short pre-trigger baseline, with $\tau_\mathrm{qp}$ estimated from the mean pulse envelope. This yields sequences of 426 samples for InHf and 120 samples for PtSi.
  \item \textbf{IQ normalization.} Both channels are divided by the global Euclidean distance from the origin to the quiescent operating point, so that the quiescent point sits at approximately unit radius.
  \item \textbf{Shuffling and splitting.} Events are randomly shuffled (fixed seed) and split into training (85\%) and validation (15\%) sets, stratified by wavelength so that every wavelength is represented in both halves. The PCA synthetic-pulse generator (Section~\ref{sec:synthetic}) is fit \emph{only} on the 85\% training subset, so neither it nor the model touches a single validation pulse during training. All resolving-power tables and figures in this paper are evaluated on the same 15\% held-out subset.
\end{enumerate}
No \emph{real data} augmentation is applied; instead, Section~\ref{sec:synthetic} describes a synthetic pulse augmentation scheme that generates training samples at continuous energies.

\section{Synthetic Pulse Generation via PCA} \label{sec:synthetic}

Calibration datasets like those used here provide $\sim$$10^4$ pulses per wavelength at discrete laser energies, with gaps of hundreds of meV between adjacent calibration points.
Training on this raw set leaves the model blind to intermediate energies and prone to overfitting on the five or ten discrete label values.
We therefore build a PCA-based generator that learns a low dimensional representation of the calibration pulses and samples new pulses at \emph{continuous} energies between calibration wavelengths.

\subsection{Streaming PCA via Gram eigendecomposition}

Let each centered \iq{} pulse be flattened to a vector in $\mathbb{R}^D$ with $D = 2\,T_\mathrm{seq}$.
Across all wavelengths the centered training pulses form a matrix $\mathbf{X} \in \mathbb{R}^{N \times D}$, with $N = 31{,}431$ pulses on PtSi and $54{,}098$ on InHf.
A straightforward SVD of $\mathbf{X}$ with full left-singular vectors allocates an $\mathcal{O}(N^2)$ buffer in float64, roughly \SI{8}{GB} on PtSi ($D = 240$, $N \approx 31{,}000$).
Instead, we accumulate the $D \times D$ Gram matrix
\begin{equation}
  \mathbf{G} = \sum_{w=1}^{N_\mathrm{wl}} (\mathbf{X}_w - \bar{\mathbf{x}})^T (\mathbf{X}_w - \bar{\mathbf{x}})
\end{equation}
streaming over wavelengths, where $\bar{\mathbf{x}}$ is the global mean pulse and $\mathbf{X}_w$ contains all pulses from wavelength $w$.
An eigendecomposition of $\mathbf{G}$ then yields the principal components $\mathbf{V}_{1:K} \in \mathbb{R}^{D \times K}$ (the top $K$ eigenvectors by descending eigenvalue) and the explained variances.
We keep $K = 50$ components throughout.
This streaming formulation matches the full SVD to machine precision ($\lesssim 10^{-14}$ relative reconstruction error, verified against the SVD path) while keeping peak memory below \SI{1}{GB}.
At $K = 50 \ll D$ this is sufficient; randomised SVD methods such as those of \citet{Halko2011} are an attractive alternative that would scale better at larger $K$, but the eigendecomposition of a $D \times D$ Gram matrix is already cheap at the dimensions used here, and streaming the wavelength-wise contributions into $\mathbf{G}$ is simpler to implement than a randomised pass over the full $\mathbf{X}$.

\begin{figure*}[htbp]
  \centering
  \includegraphics[width=\textwidth]{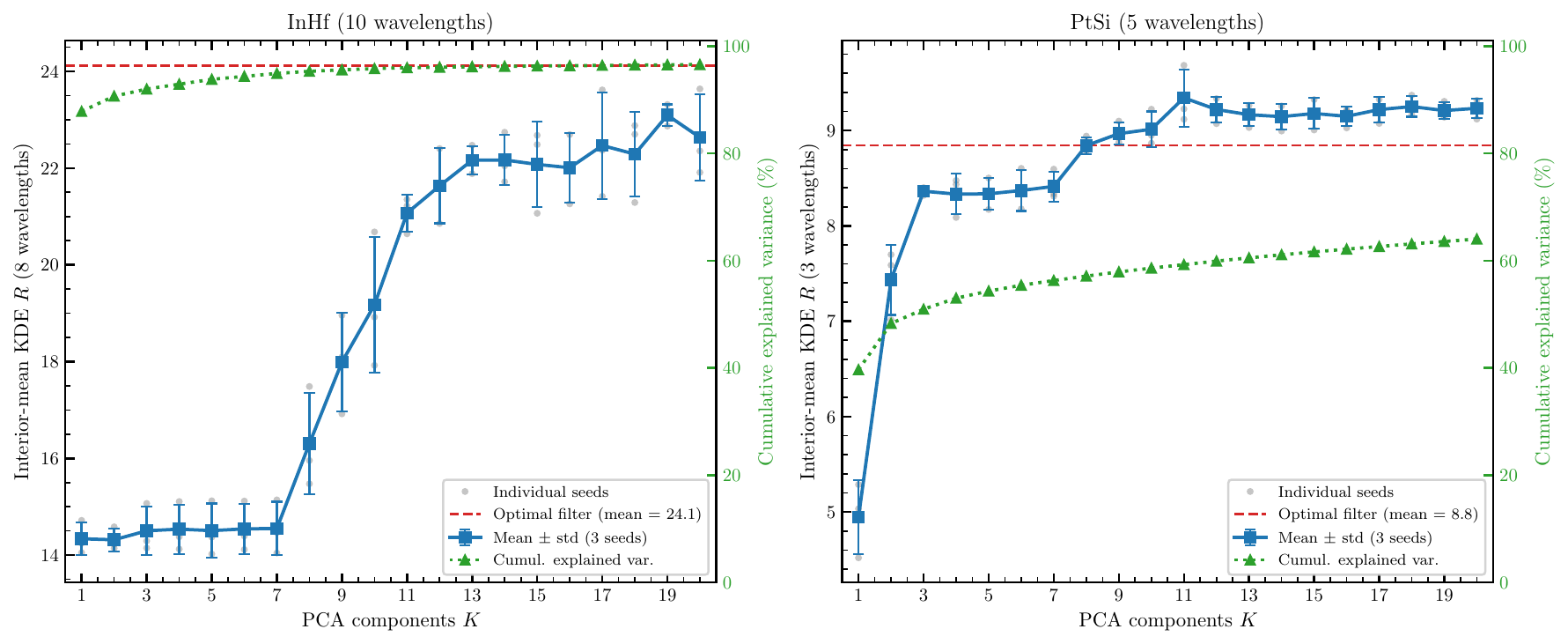}
  \caption{Interior wavelength mean KDE resolving power as a function of the number of PCA components $K$ retained by the synthetic generator, swept from $K=1$ to $K=20$. Gray dots are individual training runs (three independent seeds at each $K$); blue squares with error bars are the seed mean $\pm$ one standard deviation. The red dashed line is the interior mean optimal filter resolving power from the calibration datasets of \citet{Zobrist2022} (InHf) and \citet{Zobrist2019} (PtSi, with our reanalysis at 920 and \SI{1120}{nm}, Table~\ref{tab:results_ptsi}). The green dotted curve (right axis) is the cumulative fraction of pulse variance captured by the first $K$ components. All runs use the 25 epoch schedule, which underestimates the final InHf resolving power (Section~\ref{sec:training}). Left: InHf stays at $R\approx 14.5$ through $K=7$, rises over $K=8$--11, and levels off at $R\approx 22$--23 for $K \geq 13$. Right: PtSi reaches 8.4 at $K=3$, steps up at $K=8$ and $K=11$, and levels off at 9.1--9.3, just above the optimal filter. The full results model uses $K=50$ for both datasets.}
  \label{fig:pca_R_vs_K}
\end{figure*}

Figure~\ref{fig:pca_R_vs_K} shows how model energy resolution depends on the number of retained PCA components.
We trained the final model configuration (Section~\ref{sec:results}) with $K = 1, 2, \dots, 20$ on both datasets, repeating each run with three independent seeds, and report the mean interior wavelength KDE resolving power with one-standard-deviation error bars.
The interior wavelengths leave out the shortest and longest line of each dataset (313--\SI{1110}{nm} on InHf, 920--\SI{1120}{nm} on PtSi).
To keep the 120 runs affordable the sweep uses the 25 epoch schedule on both datasets. That schedule is converged on PtSi but not on InHf (Section~\ref{sec:training}), so the InHf curve shows how $R$ depends on $K$ rather than its final value.
The first PC alone captures $88\%$ of the InHf pulse variance but only $40\%$ of the PtSi variance. The PtSi variance is spread over many more components (the first 20 capture $64\%$ of it, against $97\%$ on InHf), as expected from its lower signal to noise ratio.
On both datasets the last gains in $R$ come from components that carry little of the variance.
For InHf, $R$ stays at 14.3--14.6 from $K=1$ to $K=7$, although these components already capture $95\%$ of the variance.
It rises over $K=8$--11 and levels off at 22.0--23.1 for $K=13$--20. Components 8 through 13 add about $1\%$ of the variance and raise $R$ from 14.6 to 22.2.
For PtSi, $R$ rises from 5.0 at $K=1$ to 8.4 at $K=3$ and stays flat through $K=7$. It steps up at $K=8$, where it reaches the interior optimal filter mean of 8.8, and again at $K=11$, then levels off at 9.1--9.3 for $K=12$--20.
The spread between seeds reaches 1.4 on InHf and 0.4 on PtSi, so we compare seed means throughout.
Both curves have leveled off by $K=13$. With the same 25 epoch schedule, $K=50$ gives interior means of $23.4 \pm 1.1$ on InHf and $9.13 \pm 0.12$ on PtSi, consistent with these plateaus. The generator cost is negligible at training time, so we keep $K=50$ for the final results runs.

\subsection{Energy interpolated sampling with PCHIP covariance}

For each calibration wavelength $w$ with energy $E_w$, we compute the per-wavelength PC-coefficient mean $\boldsymbol{\mu}_w \in \mathbb{R}^K$ and covariance $\mathbf{\Sigma}_w \in \mathbb{R}^{K \times K}$ from the projection of its pulses onto the top $K$ components.
To generate a synthetic pulse at an arbitrary energy $E$, we interpolate these statistics across the five (PtSi) or ten (InHf) calibration energies:
\begin{align}
  \boldsymbol{\mu}(E) &= \mathrm{PCHIP}(E_w \to \boldsymbol{\mu}_w)(E), \\
  \mathbf{\Sigma}(E) &= \mathrm{PCHIP}(E_w \to \mathbf{\Sigma}_w)(E).
\end{align}
Both use a shape preserving interpolator. A cubic spline applied to $\mathbf{\Sigma}_w$ can overshoot between knots, producing energies at which the interpolated covariance has smaller eigenvalues than at either calibration neighbor.
Under Gaussian NLL, the model then exploits this artifact by pulling ambiguous pulses toward the low-variance attractor (an effect we observed empirically on PtSi, where predictions piled up near the interior 980\,nm calibration energy).
A cubic spline applied to $\boldsymbol{\mu}_w$ can likewise overshoot across a wide gap between lines, where synthetic pulses would then stop tracking their energy.
The piecewise-cubic Hermite interpolating polynomial (PCHIP) of \citet{FritschCarlson1980} is monotone between adjacent knots by construction, which suppresses this overshoot.
Energies are drawn uniformly from 10\% below the lowest calibration energy to 10\% above the highest, so the model sees training pulses on both sides of every line.

Given $\boldsymbol{\mu}(E)$ and a Cholesky factor $\mathbf{L}(E)$ of $\mathbf{\Sigma}(E)$, a synthetic pulse is then
\begin{equation}
  \mathbf{p}_\mathrm{synth} = \bar{\mathbf{x}} + \mathbf{V}_{1:K}\big(\boldsymbol{\mu}(E) + \mathbf{L}(E)\,\mathbf{z}\big) + \boldsymbol{\eta},
\end{equation}
where $\mathbf{z} \sim \mathcal{N}(0, \mathbf{I}_K)$ is a fresh Gaussian draw and $\boldsymbol{\eta}$ is a reconstruction residual sample that fills the subspace orthogonal to $\mathbf{V}_{1:K}$.
We sample $\boldsymbol{\eta}$ by drawing from the empirical cross-spectral density (CSD) of the real pulse residuals and projecting it onto the orthogonal complement of the retained PCA subspace, so the full synthetic pulse reproduces both the PC-subspace statistics and the measured baseline noise colour.

\subsection{Synthetic vs real}

\begin{figure*}[htbp]
  \centering
  \includegraphics[width=\textwidth]{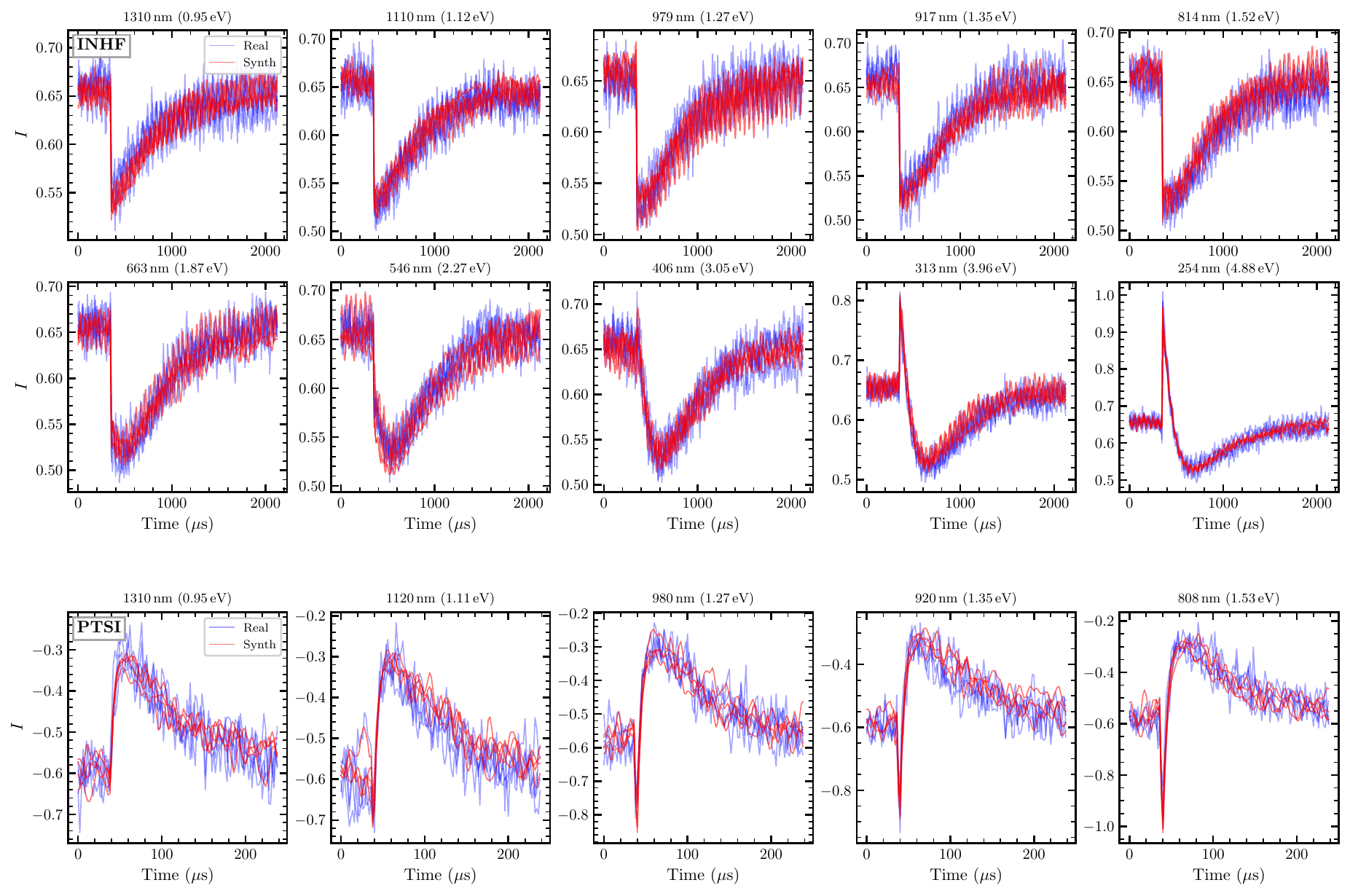}
  \caption{Synthetic vs real $I$-channel pulse comparison at every calibration wavelength for both detectors. Blue: four real pulses at the calibration energy; red: four synthetic pulses generated at the same energy from the PCA + PCHIP generator. The synthetic pulses reproduce the mean shape, amplitude, decay timescale, and baseline noise colour of the real data. The generator also samples at arbitrary intermediate energies (not shown), which is what enables continuous energy training on an otherwise discrete label calibration set.}
  \label{fig:synth_vs_real}
\end{figure*}

Figure~\ref{fig:synth_vs_real} overlays four real and four synthetic $I$-channel traces at every calibration wavelength for both datasets.
The generator captures the energy dependent pulse amplitude, decay timescale, and baseline noise spectrum at each wavelength, and extrapolates smoothly to intermediate energies.
Training uses $N_\mathrm{synth} = 120{,}000$ synthetic pulses per epoch, freshly resampled each epoch, with validation still performed on held-out \emph{real} pulses.

\section{Results} \label{sec:results}

\subsection{Training} \label{sec:training}

\begin{figure*}[htbp]
  \centering
  \includegraphics[width=\textwidth]{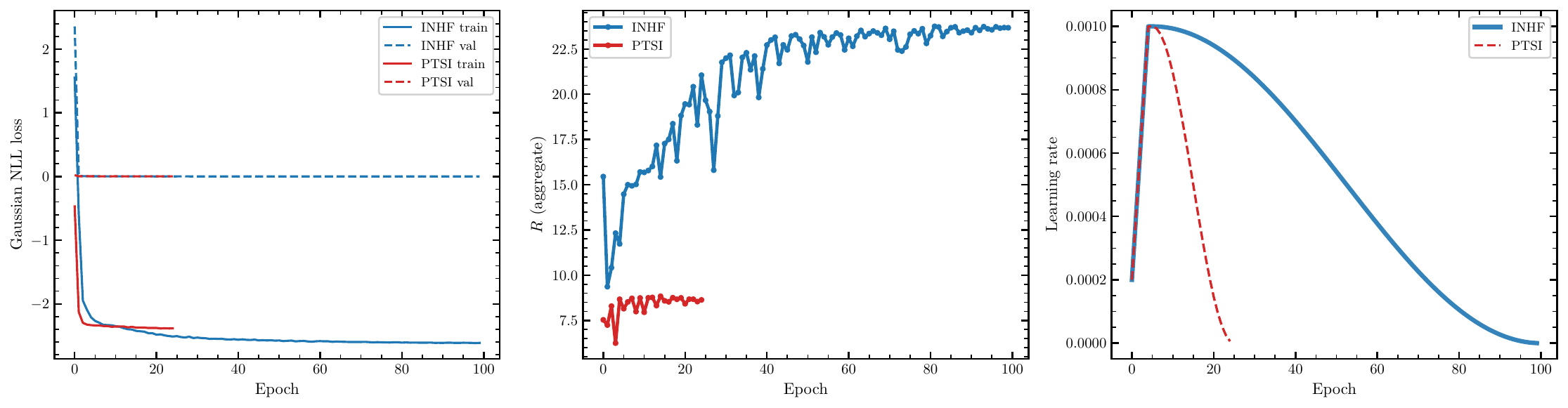}
  \caption{Training progress for both datasets (seed 42, with 25 epochs on PtSi and 100 on InHf). Left: Gaussian NLL loss on the training set (solid) and mean squared error on the held-out validation set (dashed, in eV$^2$). Middle: aggregate validation resolving power in the Gaussian approximation, $E/(2.355\,\mathrm{RMSE})$ averaged over lines, which also penalizes per-line bias and so sits below the KDE values of Tables~\ref{tab:results_inhf} and~\ref{tab:results_ptsi}. Right: cosine learning-rate schedule with five-epoch warmup. The Gaussian NLL loss looks fully converged by epoch $\sim$3 because once the predicted variance settles into the $\log\hat\sigma^2 \in [-6, 4]$ clip floor on most pulses (typically by epoch 3--5; see Section~\ref{sec:head}) the loss is dominated by an essentially constant $\log\hat\sigma^2$ term while the smaller $(E - \hat\mu)^2/\hat\sigma^2$ contribution continues to shrink. The validation $R$ is set entirely by the residuals of $\hat\mu$ and so keeps improving through epoch $\sim$70 on InHf even after the loss curve has visually flattened.}
  \label{fig:training}
\end{figure*}

\venom{} was trained with AdamW (weight decay $10^{-4}$), a cosine learning-rate schedule with five-epoch linear warmup to peak $10^{-3}$, and Gaussian NLL (Equation~\ref{eq:loss}) on 120{,}000 synthetic pulses per epoch, for 25 epochs on PtSi and 100 on InHf.
Figure~\ref{fig:training} shows the loss, aggregate resolving power, and LR schedule for both dataset runs, which share an otherwise identical recipe.
PtSi converges within the first few epochs, and a 100 epoch PtSi run gives the same resolving power. The InHf validation $R$ keeps improving to about epoch 70, so InHf trains for 100 epochs.

\textbf{Which checkpoint each table reports.}
The per-wavelength results of Tables~\ref{tab:results_inhf} and~\ref{tab:results_ptsi}, and the aggregate figures quoted in the abstract and conclusions, are means over three production checkpoints trained with seeds 42, 43 and 44.
The figures, the quantization study, and every other row that reports one seed use seed 42.
The scatter of mean $R$ between seeds is small on both detectors (10.28--10.35 on PtSi, 26.2--26.6 on InHf), so cross-table differences below about half a unit carry no information.
Comparisons within any one table are always between runs that share a seed protocol.

\subsection{Energy Resolution --- InHf} \label{sec:resolution_inhf}

Table~\ref{tab:results_inhf} lists the per-wavelength KDE resolving power of \venom{} and of the two-component optimal filter for the InHf dataset; Figure~\ref{fig:predicted_energies} (top block) shows the corresponding prediction distribution histograms.
The prediction distributions are close to Gaussian: the KDE-to-Gaussian $R$ ratio is $\sim$1.0 at every line except \SI{313}{nm}, indicating near-Gaussian residuals with only mild outlier contamination.

\begin{table}[t]
  \centering
  \caption{InHf per-wavelength KDE resolving power on the held-out validation set ($N_\mathrm{val}=9{,}470$ pulses, $15\%$ stratified hold-out, never seen during training or by the PCA synthetic pulse generator). $R_\textsc{Venom}$ is computed from the KDE FWHM of the \venom{} predictions on this subset, as the mean and standard deviation over three training seeds. $R_\mathrm{OF}$ is the published per-wavelength optimal filter resolving power from \citet{Zobrist2022}, given there to two significant figures and computed by the \texttt{mkidcalculator} pipeline (\texttt{filter\_pulses(template\_mask=True)}: a separate OF template built from each wavelength's calibration pulses) on the full dataset with that analysis's pulse cuts; restricting it to the same val subset would require Zobrist's per-wavelength filter outputs aligned to our split, which the public data files do not preserve. Mean $R_\textsc{Venom} = 26.4$, $R_\mathrm{OF} = 24.2$ ($+9\%$).}
  \label{tab:results_inhf}
  \begin{tabular}{lcccr}
    \toprule
    $\lambda$ (nm) & $E$ (eV) & $R_\textsc{Venom}$ & $R_\mathrm{OF}$ & $\Delta$ \\
    \midrule
    254  & 4.881 & $38.6 \pm 0.2$ & 35 & $+10\%$ \\
    313  & 3.961 & $47.4 \pm 1.3$ & 32 & $+48\%$ \\
    406  & 3.055 & $34.3 \pm 0.6$ & 33 & $+4\%$  \\
    546  & 2.271 & $28.3 \pm 0.2$ & 27 & $+5\%$  \\
    663  & 1.870 & $27.0 \pm 0.9$ & 25 & $+8\%$  \\
    814  & 1.524 & $21.1 \pm 0.5$ & 20 & $+6\%$  \\
    917  & 1.352 & $19.5 \pm 0.6$ & 20 & $-3\%$  \\
    979  & 1.267 & $18.9 \pm 1.1$ & 20 & $-5\%$  \\
    1110 & 1.117 & $14.5 \pm 0.3$ & 16 & $-9\%$  \\
    1310 & 0.946 & $13.8 \pm 0.5$ & 14 & $-1\%$  \\
    \bottomrule
  \end{tabular}
\end{table}

\venom{} leads the per-wavelength \texttt{mkidcalculator} optimal filter baseline at six of ten wavelengths and is within $5\%$ of it at three more, with the largest gains at the short wavelength edge (\SI{254}{nm}: $+10\%$, \SI{313}{nm}: $+48\%$) and at mid-visible (\SI{663}{nm}: $+8\%$), where the OF's energy independent template is furthest from optimal.
\venom{} trails the filter clearly only at \SI{1110}{nm} ($-9\%$).
The \SI{313}{nm} value needs a caveat. There the predictions form a narrow core on wider wings, so the KDE $R$ of 46--49 (three seeds) is far above the 23--27 implied by their RMS scatter. \venom{} still tracks the filter's energy estimate pulse by pulse at that line (correlation 0.60--0.76), so the core is not an artifact of the label, but its KDE $R$ is poorly determined (bootstrap standard deviation of 5 or more). More generally, one line's KDE $R$ carries about $10\%$ statistical uncertainty at these sample sizes.
The gap at \SI{1110}{nm} reflects the difficulty of matching a separately tuned per-wavelength OF with a single model that must generalize across the full 254--\SI{1310}{nm} range.
Overall, a single 3{,}314-parameter model trained with no per-wavelength tuning matches the published per-wavelength-template OF on aggregate KDE $R$ and exceeds it by $+9\%$ in the mean, while providing a uniform FPGA-deployable pipeline that does not require rebuilding template banks each time the detector or readout changes.
The comparison also favors the filter. Its template bank holds a separate template for each laser line, selected by knowing the photon's wavelength, which is the quantity being measured. Such a bank cannot be applied to photons of unknown energy, so even matching it with one model is useful.

\subsection{Energy Resolution --- PtSi} \label{sec:resolution_ptsi}

\begin{figure*}[htbp]
  \centering
  \includegraphics[width=\textwidth]{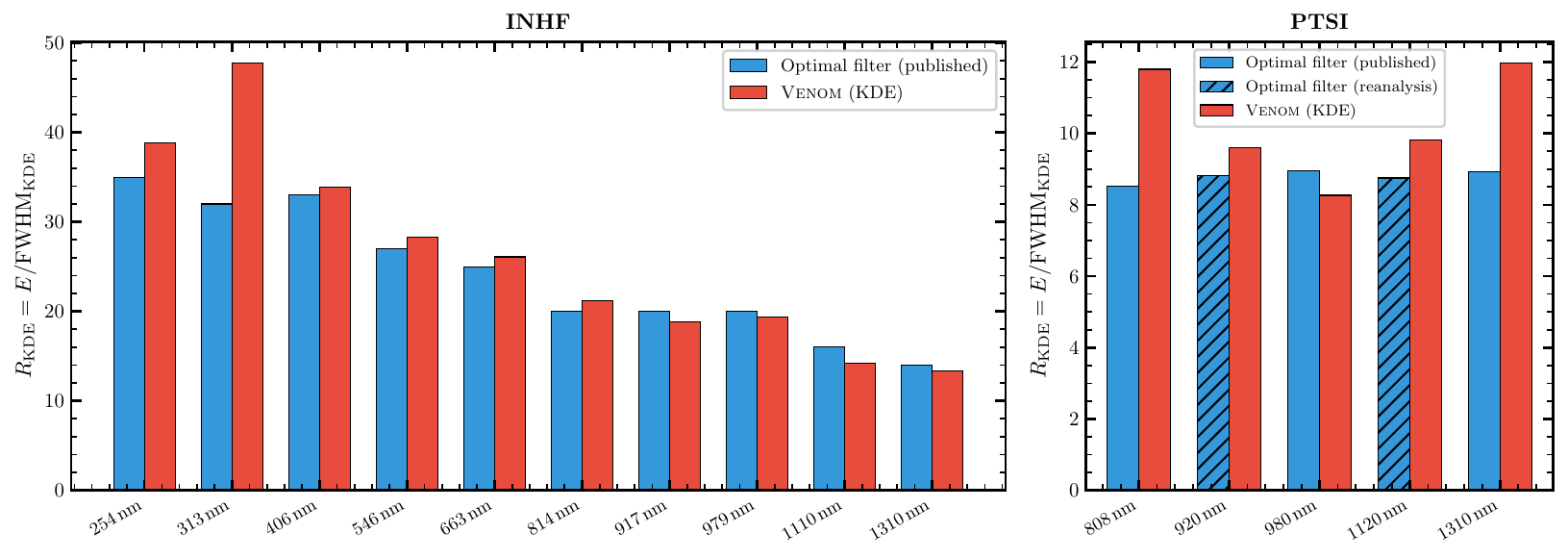}
  \caption{Per-wavelength KDE resolving power for \venom{} (red, seed 42 checkpoints) vs.\ the two-component (phase and dissipation) optimal filter (blue). Left: InHf (10 wavelengths, 254--\SI{1310}{nm}). Right: PtSi (5 wavelengths, 808--\SI{1310}{nm}). The hatched filter bars at 920 and \SI{1120}{nm}, where no value was published, come from our reanalysis (Table~\ref{tab:results_ptsi}). Both panels use the same 3{,}314-parameter model and identical training recipe.}
  \label{fig:comparison}
\end{figure*}

Table~\ref{tab:results_ptsi} and Figure~\ref{fig:comparison} present the PtSi results.
\venom{} reaches a mean KDE $R$ of 10.3 vs.\ 8.8 for the shared template optimal filter.
At the interior wavelengths (920, 980, \SI{1120}{nm}) the two agree within $11\%$ (9.3, 8.3 and 9.7 against 8.8, 9.0 and 8.8, a mean of 9.1 against 8.8), and the excess at the two edge wavelengths depends on a training choice, as discussed below.

\begin{table}[t]
  \centering
  \caption{PtSi per-wavelength KDE resolving power on the held-out validation set ($N_\mathrm{val}=5{,}569$, $15\%$ stratified hold-out, never seen during training or by the PCA synthetic generator). $R_\textsc{Venom}$ is computed from the KDE FWHM of the \venom{} predictions on this subset, as the mean and standard deviation over three training seeds. $R_\mathrm{OF}$ is the resolving power of the published analysis of these data, which used one shared (920~nm) two-component optimal filter template \citep{Zobrist2019} and reported $R$ at 808, 980 and 1310~nm. $^\ddagger$Not published. From our reanalysis of the traces with the published recipe (920~nm filter, interpolating spline energy calibration through zero), which reproduces the other three values within 2\%. With a smooth quadratic calibration instead, the filter's $R$ changes by 2--9\% at the interior lines and by up to 27\% at 808~nm, a systematic that applies to the published values as well. Mean $R_\textsc{Venom} = 10.3$, $R_\mathrm{OF} = 8.8$.}
  \label{tab:results_ptsi}
  \begin{tabular}{lcccr}
    \toprule
    $\lambda$ (nm) & $E$ (eV) & $R_\textsc{Venom}$ & $R_\mathrm{OF}$ & $\Delta$ \\
    \midrule
    808  & 1.534 & $12.0 \pm 0.3$ & 8.5 & $+41\%$ \\
    920  & 1.348 & $9.3 \pm 0.6$  & 8.8$^\ddagger$ & $+6\%$  \\
    980  & 1.265 & $8.3 \pm 0.1$  & 9.0 & $-7\%$  \\
    1120 & 1.107 & $9.7 \pm 0.3$  & 8.8$^\ddagger$ & $+11\%$ \\
    1310 & 0.946 & $12.1 \pm 0.3$ & 8.9 & $+36\%$ \\
    \bottomrule
  \end{tabular}
\end{table}

The edge wavelengths depend on how far past them the synthetic energies extend.
Widening that range from the default 10\% to 20\% and 30\% (seed 42) lowers $R$ at \SI{808}{nm} from 11.8 to 9.4 and 7.7, and at \SI{1310}{nm} from 12.0 to 9.8 and 9.7, while the three interior wavelengths stay within 8.2--10.4.
A narrow range truncates the energy prior just outside an edge line, and the regression then compresses its predictions there. A wide range instead trains on PCHIP statistics extrapolated far past the last calibration line.
No setting avoids both effects, so we regard the PtSi edge values as upper limits and the interior wavelengths as the fair comparison.
On InHf the same test lowers $R$ at \SI{1310}{nm} from 13.3 to 12.2 and 11.6 and changes it at \SI{254}{nm} from 38.8 to 39.2 and 44.8, so the InHf edge values carry a systematic of similar size.

This agrees with the expectation that the PtSi energy resolution is limited by athermal phonon escape to near the values seen with the optimal filter~\citep{Zobrist2022}.

\subsection{Comparison with Learned Baselines} \label{sec:baselines}

We ran three matched recipe ablations alongside the published \venom{} configuration to test whether the \ssm{} backbone is doing real work.
The first is a hand-feature MLP that takes [peak height, peak time, 10--90 rise time, 90--10 decay time, integral, pre-trigger RMS, FWHM, half-decay] as input to a 3{,}290-parameter feedforward network.
The second is a non-selective S4D ablation that replaces Mamba's input dependent $\mathbf{B}$, $\mathbf{C}$, and $\Delta$ with learnable constants while keeping the rest of the architecture.
The third is a causal TCN (Temporal Convolutional Network) that swaps the \ssm{} backbone for four causal dilated depthwise + pointwise conv blocks (kernel 3, dilations $[1, 2, 4, 8]$) at matched parameter count.
All four models share the same training (Gaussian NLL, AdamW with weight decay $10^{-4}$, cosine schedule with 5-epoch warmup, 25 epochs on PtSi and 100 on InHf for the \iq{}-stream models and 100 for the MLP, $120{,}000$ PCA-synthetic pulses per epoch for the \iq{}-stream models) and the same $15\%$ stratified hold-out used elsewhere in this paper.
Table~\ref{tab:ablations} reports the means over three seeds.

\begin{table}[t]
  \centering
  \caption{Mean KDE resolving power on the held-out validation set ($N_\mathrm{val}^\mathrm{PtSi}=5{,}569$, $N_\mathrm{val}^\mathrm{InHf}=9{,}470$) for the four models compared in Section~\ref{sec:baselines}, averaged over three seeds. All four share training recipe and val split. $^\dagger$The MLP's PtSi predictions collapse onto a spike narrower than the KDE can resolve, so its PtSi entry is the RMS-based $E/(2.355\,\mathrm{RMSE})$. Checkpoints for all runs are shipped in \texttt{opt/paper/}.}
  \label{tab:ablations}
  \begin{tabular}{lrrr}
    \toprule
    Model               & Params & PtSi $\bar{R}$ & InHf $\bar{R}$ \\
    \midrule
    Optimal Filter      & ---    & 8.8            & 24.2 \\
    Hand-feature MLP    & 3{,}290 & 6.0$^\dagger$ & 2.5 \\
    Non-selective S4D   & 2{,}818 & 10.2           & 26.1 \\
    Causal TCN          & 3{,}234 & 10.3           & 27.3 \\
    \venom{}            & 3{,}314 & 10.3           & 26.4 \\
    \bottomrule
  \end{tabular}
\end{table}

The hand-feature MLP falls short of the optimal filter even on the narrow-band PtSi calibration (5 wavelengths from 0.95 to \SI{1.53}{eV}) but collapses to $R < 5$ across the wide-band InHf range (10 wavelengths from 0.95 to \SI{4.88}{eV}). The failure is structural, not a capacity limit. \venom{} reaches $\bar R = 26.4$ on the same data at essentially the same parameter budget (3{,}314 vs 3{,}290), so what matters is keeping the full pulse shape rather than projecting it down to eight summary scalars (peak height, rise/decay times, FWHM, integral, pre-trigger RMS, half-decay), which discard the pulse shape information needed to discriminate photon energies across a $5\times$ range.
The non-selective S4D ablation matches \venom{} within the scatter between seeds on both detectors (PtSi 10.2 vs.\ 10.3; InHf 26.1 vs.\ 26.4), so on these data the input dependent $\mathbf{B}$, $\mathbf{C}$, and $\Delta$ projections in the Mamba block add no measurable resolution.
The causal TCN at matched parameter count ties \venom{} on PtSi and is $3\%$ ahead on InHf (27.3 vs.\ 26.4).
Both architectures admit the same $O(1)$-per-sample recurrent inference required for FPGA deployment.
We retain the \ssm{} design for two practical reasons.
First, the recurrent state per pulse is smaller: the \ssm{} holds $d_\mathrm{model} \times d_\mathrm{state} = 128$ floats per layer while the dilated TCN at kernel 3 with dilations $[1, 2, 4, 8]$ needs $(k - 1) \times d_\mathrm{max} \times d_\mathrm{model} = 256$ floats per layer's ring buffer.
Second, Mamba's input dependent $\Delta$ adapts the effective integration timestep across pulse regimes (quiescent baseline, rising edge, decay tail), which we expect will help generalization to detector geometries outside the training calibration.
On energy resolution alone, the dilated TCN is a serviceable alternative.

\subsection{Architecture Ablations} \label{sec:arch_sweep}

Table~\ref{tab:arch_sweep} reports a one-knob-at-a-time sweep around the published winner ($d_\mathrm{model}=16$, $d_\mathrm{state}=8$, $n_\mathrm{layers}=1$, $d_\mathrm{conv}=4$, expand $=1$, attention pooling, gated branch on) on both calibration datasets, holding training recipe and validation split fixed.
Each variant is one seed 42 run, compared with the seed 42 winner, so differences below about $0.6$ in $R$ carry no information.
Attention pooling beats mean and last-token pooling on both detectors, and max pooling ties it.
The mean-pool deficit is $0.5$ in $R$ on PtSi and $1.5$ in $R$ on InHf; the last-token pool is $0.5$ in $R$ worse on PtSi and $5.7$ in $R$ worse on InHf.
The gap widens on InHf because the longer 426-sample window (vs.\ PtSi's 120) gives more time for noise to accumulate at non-trigger positions, exactly the regime where attention's learned weighting outperforms a uniform average.
The Section~\ref{sec:head} claim that the model uses time localized features is supported on both detectors and especially load-bearing on the longer-window dataset.
The $\mathrm{silu}(z)$ gated branch is neutral on both detectors (mean $R$ changes by $0.0$ on PtSi and $+0.2$ on InHf with the gate off). We keep it to stay with the standard Mamba block, although dropping it would save 256 parameters.
For the SSM dimensions, the $(d_\mathrm{model}, d_\mathrm{state}, n_\mathrm{layers}, d_\mathrm{conv})$ winner sits at or near a Pareto point on PtSi, where every variant lies within $0.3$ in $R$ of it, but not on InHf, where added capacity still pays.
Doubling $d_\mathrm{model}$ changes $R$ by $-0.2$ on PtSi and $+2.1$ on InHf for $3.5\times$ parameters; $n_\mathrm{layers}=2$ by $-0.1$ on PtSi and $+0.8$ on InHf for $+47\%$ parameters; doubling expand by $+0.1$ on PtSi and $+1.6$ on InHf at $+60\%$ parameters; and kernels of 1, 2, and 8 give $R$ within $0.3$ of $d_\mathrm{conv}=4$ on PtSi and within $0.8$ on InHf.
$d_\mathrm{conv}=4$ is therefore a reasonable default for both datasets, matching the Section~\ref{sec:backbone} motivation that the depthwise conv should span the rising edge.
Halving $d_\mathrm{state}$ to 4 holds $R$ within $0.1$ of the winner on both datasets at 200 fewer parameters, while doubling it to 16 gains $1.2$ on InHf.
Halving $d_\mathrm{model}$ to 8 (1{,}018 parameters) costs $1.1$ on InHf and nothing on PtSi, an option where the FPGA budget is tight, and the InHf gains from larger models are worth picking up in a follow-up retrain.

\begin{table}[t]
  \centering
  \caption{One-knob-at-a-time architecture sweep on both datasets. All rows share training recipe and the same $15\%$ stratified hold-outs used in Tables~\ref{tab:results_inhf} and~\ref{tab:results_ptsi}. $\bar{R}$ is the unweighted mean KDE $R$ across the dataset's wavelengths. Each row varies one knob from the published winner, and every row, including the winner, is the seed 42 run. Over three seeds the winner gives $10.31 \pm 0.04$ on PtSi and $26.36 \pm 0.23$ on InHf. Summed over runs, training took \SI{129}{min} on PtSi and \SI{879}{min} on InHf on a single RTX~5090, three runs at a time; checkpoints and per-wavelength tables are in \texttt{opt/paper/arch/}.}
  \label{tab:arch_sweep}
  \begin{tabular}{lrrr}
    \toprule
    Configuration & Params & PtSi $\bar{R}$ & InHf $\bar{R}$ \\
    \midrule
    \textbf{Winner} & 3{,}314 & \textbf{10.3} & \textbf{26.2} \\
    \midrule
    \multicolumn{4}{l}{\emph{Pooling}} \\
    \quad mean        & 3{,}298  & 9.8  & 24.7 \\
    \quad max         & 3{,}298  & 10.4 & 26.0 \\
    \quad last-token  & 3{,}298  & 9.8  & 20.4 \\
    \midrule
    \multicolumn{4}{l}{\emph{Gated branch}} \\
    \quad off         & 3{,}058  & 10.3 & 26.4 \\
    \midrule
    \multicolumn{4}{l}{\emph{SSM dimensions}} \\
    \quad $d_\mathrm{model}=8$  & 1{,}018  & 10.5 & 25.1 \\
    \quad $d_\mathrm{model}=32$ & 11{,}746 & 10.1 & 28.3 \\
    \quad $d_\mathrm{state}=4$  & 3{,}122  & 10.3 & 26.1 \\
    \quad $d_\mathrm{state}=16$ & 3{,}698  & 10.3 & 27.3 \\
    \quad $n_\mathrm{layers}=2$ & 4{,}866  & 10.2 & 27.0 \\
    \quad $d_\mathrm{conv}=1$   & 3{,}266  & 10.5 & 26.0 \\
    \quad $d_\mathrm{conv}=2$   & 3{,}282  & 10.4 & 26.9 \\
    \quad $d_\mathrm{conv}=8$   & 3{,}378  & 10.4 & 27.0 \\
    \quad expand $=2$           & 5{,}346  & 10.4 & 27.7 \\
    \bottomrule
  \end{tabular}
\end{table}

\subsection{Analysis} \label{sec:analysis}

\begin{figure*}[htbp]
  \centering
  \includegraphics[width=\textwidth]{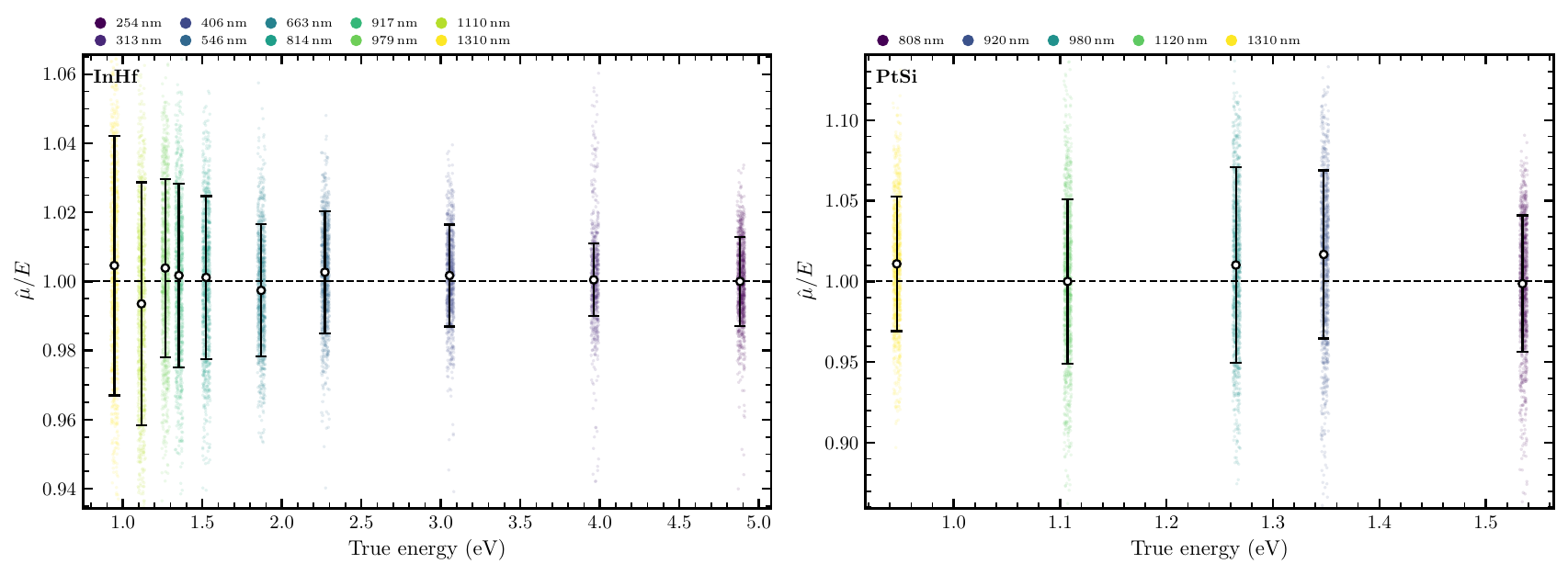}
  \caption{Ratio of predicted to true photon energy, $\hat\mu/E$, against true energy for all validation events, coloured by wavelength. Left: InHf. Right: PtSi. The dashed line at unity is a perfect estimate. Black markers give the per-wavelength median with whiskers of half-height $1/2R$, where $R$ is the KDE resolving power of Tables~\ref{tab:results_inhf} and~\ref{tab:results_ptsi}, so the whisker spans the FWHM and the panel reads directly as resolving power against energy. Point opacity is low so dense clusters saturate while isolated outliers stay faint. The horizontal scatter within each laser line is presentational: the calibration energies are discrete, and points are jittered by $\pm0.6\%$ of the plotted energy range so that each line reads as a band rather than a single column.}
  \label{fig:scatter}
\end{figure*}

\begin{figure*}[htbp]
  \centering
  \includegraphics[width=\textwidth]{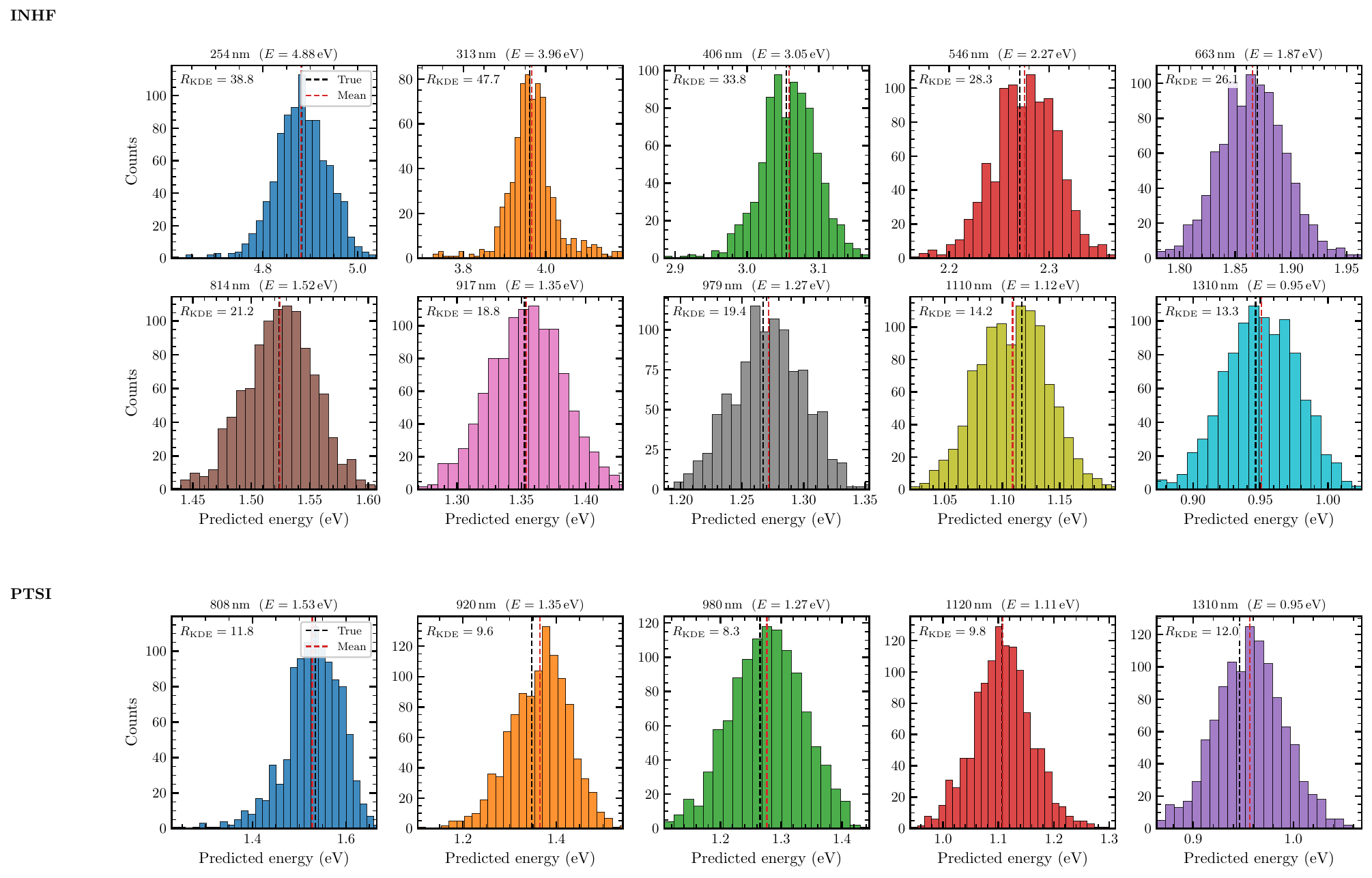}
  \caption{Per-wavelength \venom{} predicted energy distributions on the held-out validation set. Top block: InHf; bottom: PtSi. Black dashed lines mark the true photon energy $E = hc/\lambda$; red dashed lines mark the mean \venom{} prediction. KDE $R$ for each wavelength is quoted in the upper-left corner of each panel.}
  \label{fig:predicted_energies}
\end{figure*}

Figure~\ref{fig:scatter} shows the ratio of predicted to true energy against true energy for both datasets.
The dense band at unity is the bulk population, and the sparse halo above and below it is the residual outlier tail.
Both datasets sit close to unity at every calibration line, so the estimator is approximately unbiased, and the whiskers narrow towards higher energy on InHf, from $\pm3.8\%$ at \SI{1310}{nm} to $\pm1.3\%$ at \SI{254}{nm}, which is the energy dependence of $R$ read directly off the figure.
Figure~\ref{fig:predicted_energies} shows the corresponding per-wavelength prediction distribution histograms, with the true laser energy (black dashed) and the mean prediction (red dashed) marked on each panel. There is some uncertainty in the true laser wavelengths, as we are relying on the manufacturer's wavelength calibration.
The proximity of mean to true across nearly all wavelengths reflects the small systematic bias of \venom{}'s Gaussian NLL fit; the remaining residual is dominated by random scatter, which is what sets $R$.
The residual per-wavelength offsets are at most \SI{8}{meV} on InHf and \SI{17}{meV} on PtSi (at \SI{920}{nm}). They are not an outlier effect: trimming the top $5\%$ of validation pulses by $|E-\hat\mu|$ shifts each by at most \SI{8}{meV}. The offset is removable offline by a quadratic energy calibration of the predicted-vs-true centroids, exactly the post-hoc step already applied to optimal filter outputs in the published \mkid{} pipeline \citep{Zobrist2019, Zobrist2022}; this calibration is operationally free at deployment time and does not affect the resolving power. Paired with the lifted-clip variance head of Section~\ref{sec:head}, the same step also delivers a calibrated per-pulse uncertainty, with the outlier flag $|E - \hat\mu| > 3\,\hat\sigma$ on both detectors.

\subsection{Learned Feature Representations} \label{sec:features}

\begin{figure*}[htbp]
  \centering
  \includegraphics[width=\textwidth]{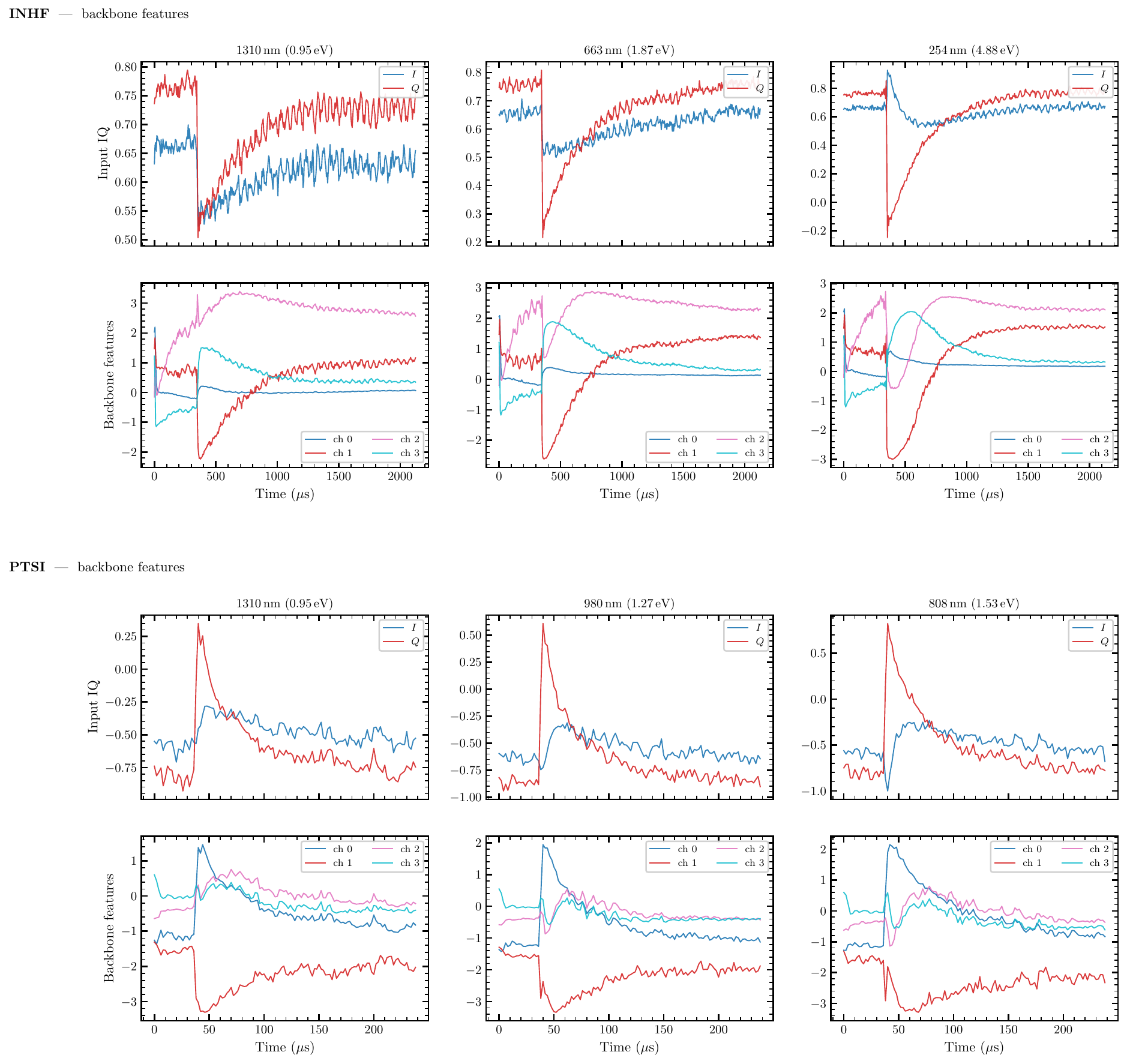}
  \caption{Top rows of each dataset block: raw normalized \iq{} traces for single photon events at three wavelengths spanning each dataset's energy range. Bottom rows: the four backbone feature channels for the same pulses. The learned features encode pulse onset, decay dynamics, and energy dependent variations that the energy head uses for regression.}
  \label{fig:features}
\end{figure*}

Figure~\ref{fig:features} shows the raw \iq{} input and four backbone feature channels for single photon events at three representative wavelengths from each dataset.
The single Mamba block transforms the two-channel \iq{} input into a 16-dimensional representation in which different feature channels clearly encode different physical aspects of the pulse: some respond sharply to the pulse onset, others track the slower quasiparticle recombination tail, and some channels carry approximately energy independent baseline information.
The interpretability of these features was not engineered but emerges from the single layer backbone, which is deep enough to learn the coordinate transform and filtering but shallow enough that individual channels retain physical identity.

\section{Path to FPGA Deployment} \label{sec:fpga}

A primary design goal of \venom{} is deployment on the MKIDGen3 RFSoC readout electronics \citep{Smith2024}, which must process thousands of resonator channels simultaneously in real time.
The current pipeline uses an optimal filter combined with a 2D quadratic coordinate transform, which \citet{Smith2024} noted is ``not feasible for an FPGA-based instrument at our target scale.''
\venom{}'s recurrent inference mode, together with the 3{,}314-parameter model size, may offer a substantially more tractable alternative.

\textbf{Computational cost.}
In recurrent mode, each ADC sample requires a single block hidden state update.
The input dependent projections cost $\dstate \times \dinner = 128$ MACs each for $\mathbf{W}_B$ and $\mathbf{W}_C$ and $\dinner \times \dinner = 256$ for $\mathbf{W}_\Delta$, or $512$ in total.
The block's expansion and output projections add $2\dmodel\dinner + \dmodel\dinner = 768$, the raw \iq{} input projection and the depthwise convolution contribute $96$, and the state update together with the output contraction over the $\dstate$ modes contributes $544$.
The total is approximately 1{,}900\,MACs per sample, alongside $\dinner \dstate = 128$ exponentials for the discretisation, which are not MACs and would be handled by a lookup table or piecewise-linear approximation in firmware.
This is within the DSP capacity of modern RFSoCs at the effective rates of both datasets (\SI{200}{kHz}--\SI{500}{kHz} here, easily extensible to the \SI{1}{MHz}--\SI{2}{MHz} raw ADC rate of MKIDGen3).
Tier~2 is excluded from this count because it runs once per photon rather than once per sample.

\textbf{Weight memory.}
The 3{,}314 parameters require \SI{6.6}{kB} at 16-bit precision and approximately \SI{4.3}{kB} under W8A32 (INT8 weights, FP32 biases and per-tensor scales; see the next paragraph).

\textbf{Streaming architecture.}
The recurrent state update (Equation~\ref{eq:state_update}) maps naturally to a streaming pipeline: at each clock cycle, the FPGA reads one \iq{} sample, updates the single layer hidden state, and writes the backbone features to a circular buffer.
When a photon trigger is detected (from a threshold on the feature magnitude or a separate trigger channel), the energy head reads the buffered features and produces an energy estimate.
The Tier~1/Tier~2 separation ensures that the energy head's MLP computation occurs only at the photon rate, not the ADC rate.

\textbf{Post-training INT8 weight quantization.}
To substantiate the W8A32 weight memory budget, we replace every weight tensor in each winner checkpoint with a symmetric per-tensor INT8 quantize/dequantize round-trip (biases and the small 1-D parameters of the SSM core stay FP32, $\sim$9\% of the parameter count) and re-evaluate per-wavelength KDE $R$ on the same held-out validation set used elsewhere in this paper.
Table~\ref{tab:ptq} reports the result.
On InHf the mean $R$ moves by $-0.50$ ($26.16 \rightarrow 25.66$), or $2\%$; on PtSi by $+0.04$ ($10.29 \rightarrow 10.33$), within seed-to-seed training variance (Table~\ref{tab:loss_ablation}).
This bounds the weight-quantization-only contribution to the deployment error budget, with full activation quantization, fixed-point accumulator width selection for the SSM state, and quantization-aware fine-tuning deferred to a planned firmware paper.


\begin{table}[t]
  \centering
  \small
  \setlength{\tabcolsep}{4pt}
  \renewcommand{\arraystretch}{1.08}
  \caption{Post-training weight quantization (W8A32: INT8 weights, FP32 activations) on the published winners. Symmetric per-tensor scaling, dequantize on load; biases and per-tensor scale factors stay FP32. The FP32 row is recomputed from the same checkpoint on the same val subset to make the $\Delta R$ a pure quantization effect, not a split-stratification artifact.}
  \label{tab:ptq}
  \begin{tabular}{lcccc}
    \toprule
    Detector & Weights@8b & Total bytes & $\bar R_{\rm FP32}$ & $\bar R_{\rm W8A32}$ ($\Delta\bar R$) \\
    \midrule
    InHf & 3,008\,B & 4,280\,B & 26.16 & 25.66 ($-0.50$) \\
    PtSi & 3,008\,B & 4,280\,B & 10.29 & 10.33 ($+0.04$) \\
    \bottomrule
  \end{tabular}
\end{table}

\textbf{Activation quantization.}
The selective mechanism produces input dependent $\mathbf{B}$, $\mathbf{C}$, and $\boldsymbol{\Delta}$ values which require careful dynamic range analysis to avoid overflow in the exponential discretization (Equation~\ref{eq:state_update}); the SSM hidden state $\mathbf{h}_i[k]$ furthermore accumulates over hundreds of samples, so the relevant fixed-point design choice is the SSM-state accumulator width rather than the weight precision.
Quantization-aware training and detailed resource utilization benchmarks for an MKIDGen3 firmware build are left for future work.

The approximate parallel-vs-recurrent equivalence (Section~\ref{sec:modes}, Appendix~\ref{app:parrec}) is the foundational property enabling this deployment path: the model is trained in parallel mode on a GPU and deployed in recurrent mode on FPGA hardware without retraining or additional calibration.

\section{Conclusions} \label{sec:conclusions}

We have presented \venom{}, a selective state space model that replaces the coordinate transform and matched filter in the \mkid{} photon energy estimation pipeline with a single learned sequence model operating directly on raw \iq{} timestream data.
The same 3{,}314-parameter model, trained with a PCA-based synthetic pulse generator that samples at continuous energies between calibration wavelengths, delivers:
\begin{itemize}
  \item A mean KDE resolving power of 26.4 on an InHf bilayer detector covering ten wavelengths from 254 to \SI{1310}{nm}, compared with 24.2 for the published per-wavelength optimal filter baseline. This is a $9\%$ aggregate improvement using a single model in place of ten templates, with \venom{} leading at six of ten wavelengths and within $5\%$ of the filter at three more.
  \item A KDE $R$ of 9.1 on average at the three interior wavelengths of a PtSi detector covering five wavelengths from 808 to \SI{1310}{nm}, matching the shared template optimal filter baseline (8.8 at those wavelengths).
\end{itemize}

Three elements carry the architecture:
(1) a single selective-\ssm{} block is sufficient to learn both the coordinate transform and the matched filtering, removing the need for custom preprocessing;
(2) Gaussian NLL training matches MSE, Huber and $\beta$-NLL in resolving power and is stable across seeds (Table~\ref{tab:loss_ablation}). The variance head in the published recipe is not a calibrated uncertainty (Section~\ref{sec:head}) and we read out only $\hat\mu$, but a lifted-clip $\beta$-NLL variant gives, at no measurable cost in $R$, a $\hat\sigma$ calibrated over the whole validation set on both detectors that downstream applications can use as a per-pulse uncertainty plus a $3\sigma$ outlier flag;
(3) PCA-plus-PCHIP synthetic data turn a small discrete energy calibration set into a continuous energy training distribution, critical when per-wavelength event counts are $\lesssim$$10^4$.

Several directions remain.
Per-pulse input dependent covariance modeling in the synthetic generator (beyond the linear PCA subspace) should narrow the remaining KDE-vs-Gaussian gap on PtSi.
Hybrid real+synthetic training may mitigate the PtSi 1310\,nm edge effect where the PCA covariance must extrapolate.
Multi-resonator conditioning where we make the \ssm{} parameters a function of per-channel calibration vectors could amortize a single trained \venom{} instance across an entire detector array.
Pile-up detection via the backbone's latent state, quantization-aware training for fixed-point FPGA deployment, and sub-sample photon-arrival-time regression are natural extensions.

The VENOM approach is likely to transfer to other detectors where non-stationary noise and pulse shape variation are present, such as Magnetic Microcalorimeters~\citep{Kempf2018}, Transition Edge Sensors~\citep{Ullom2015}, and phonon-mediated KIDs~\citep{Temples2024}, although small modifications will be needed to deal with a single coordinate (like current pulses in TES) readout. There do not appear to be any fundamental reasons this approach cannot reach the very high R${\sim}$6000 seen in X-ray TESs.

The combination of optimal filter energy resolution on two materials without per-wavelength templates, a 3{,}314-parameter footprint, and a natural streaming architecture makes \venom{} a promising candidate for next-generation \mkid{} readout. The VENOM code is available at \url{https://github.com/MazinLab/VENOM} under a BSD-3-Clause license.

\section*{Data Availability} \label{sec:data_availability}

The \venom{} code that supports the findings of this article is openly available~\cite{VENOMcode}. The \mkid{} calibration data analyzed here are the published measurements of Refs.~\cite{Zobrist2022} and~\cite{Zobrist2019}; requests for further information or data should be sent to the authors.


\begin{acknowledgments}
We thank Nicholas Zobrist for making the PtSi and InHf \mkid{} calibration datasets publicly available. This material is based upon work supported by the National Aeronautics and Space Administration under Grant No. 80NSSC19K0329 and NNX16AE98G. This work made use of the MLX framework \citep{MLX2023} for model development on Apple Silicon and PyTorch on CUDA. Claude was used extensively for this development.

\end{acknowledgments}


\appendix

\section{Zero-Order Hold Discretization} \label{app:zoh}

The continuous-time state space model (Equations~\ref{eq:ssm_cont_h}--\ref{eq:ssm_cont_y}) is discretized for sampled data using the zero-order hold (ZOH) assumption: the input $u(t)$ is held constant at $u[k]$ over the interval $[k\Delta, (k\!+\!1)\Delta)$.
Integrating Equation~\eqref{eq:ssm_cont_h} over this interval gives
\begin{equation}
  \mathbf{h}((k\!+\!1)\Delta) = e^{\mathbf{A}\Delta}\,\mathbf{h}(k\Delta) + \left(\int_0^\Delta e^{\mathbf{A}\tau}\,\mathrm{d}\tau\right)\mathbf{B}\,u[k].
\end{equation}
Evaluating the integral yields the discrete-time matrices:
\begin{align}
  \bar{\mathbf{A}} &= e^{\mathbf{A}\Delta}, \\
  \bar{\mathbf{B}} &= \mathbf{A}^{-1}(e^{\mathbf{A}\Delta} - \mathbf{I})\,\mathbf{B}.
\end{align}
For a diagonal state matrix $\mathbf{A} = \diag(a_1, \ldots, a_N)$ with $a_n < 0$, these reduce to element-wise operations:
\begin{align}
  \bar{a}_n &= e^{a_n \Delta}, \\
  \bar{b}_n &= \frac{e^{a_n \Delta} - 1}{a_n}\,b_n.
\end{align}
The diagonal structure is essential for computational efficiency: no matrix inverse or matrix exponential is needed, only scalar exponentials and divisions.
In the selective \ssm{}, the timestep $\Delta$ becomes input dependent ($\Delta[k]$), so these quantities are recomputed at each sample.

\section{Parallel-vs-Recurrent Equivalence Distribution} \label{app:parrec}

The two \ssm{} code paths described in Section~\ref{sec:modes} should compute the same function in exact arithmetic; in practice they accumulate FP32 round-off differently, and the recurrent mode is what runs on streaming hardware. Table~\ref{tab:parrec} reports the per-pulse energy disagreement $|\hat\mu_\| - \hat\mu_{\rm rec}|$ on the full validation set for the two winner models, both in absolute $\mu$eV and normalised to the parallel mode statistical resolution $\sigma_\|$. The maximum disagreement is well under $\sigma_\|$ on both datasets, and the per-wavelength KDE-$R$ shift on recurrent deployment is bounded by $|\Delta R| \le 0.03$ across all fifteen wavelengths.

\begin{table}[t]
\centering
\small
\setlength{\tabcolsep}{4pt}
\renewcommand{\arraystretch}{1.08}
\caption{Parallel-vs-recurrent energy disagreement distribution on the full validation set for the two winner models. $\sigma_\|$ is the pooled standard deviation of parallel mode residuals (a measure of per-pulse statistical resolution). Percentiles of the per-pulse disagreement $|\hat{\mu}_\|-\hat{\mu}_{\rm rec}|$ are reported in absolute $\mu$eV and normalized to $\sigma_\|$. $\Delta R_{\max}$ is the largest per-wavelength KDE-$R$ shift from parallel to recurrent deployment; with the Hillis--Steele prefix scan, every wavelength matches to within $|\Delta R|\le 0.03$.}
\label{tab:parrec}
\begin{tabular}{lcc}
\toprule
& \textbf{InHf} & \textbf{PtSi} \\
 & (seq.~len.~426) & (seq.~len.~120) \\
\midrule
$\sigma_\|$ (meV) & 38 & 57 \\
\addlinespace
\multicolumn{3}{l}{$|\hat\mu_\| - \hat\mu_{\rm rec}|$ ($\mu$eV):} \\
\quad median      & 0.12 & 0.12 \\
\quad 95th pct    & 0.48 & 0.24 \\
\quad 99th pct    & 0.95 & 0.36 \\
\quad 99.9th pct  & 1.43 & 0.48 \\
\quad max         & 1.91 & 0.60 \\
\addlinespace
\multicolumn{3}{l}{in units of $\sigma_\|$:} \\
\quad median      & $3\times10^{-6}$ & $2\times10^{-6}$ \\
\quad 99th pct    & $3\times10^{-5}$ & $6\times10^{-6}$ \\
\quad max         & $5\times10^{-5}$ & $1\times10^{-5}$ \\
\addlinespace
$\Delta R_{\max}$ & $+0.03$ (\SI{313}{nm}) & $0.00$ \\
\bottomrule
\end{tabular}
\end{table}

\bibliography{references}

\begin{thebibliography}{29}%
\makeatletter
\providecommand \@ifxundefined [1]{%
 \@ifx{#1\undefined}
}%
\providecommand \@ifnum [1]{%
 \ifnum #1\expandafter \@firstoftwo
 \else \expandafter \@secondoftwo
 \fi
}%
\providecommand \@ifx [1]{%
 \ifx #1\expandafter \@firstoftwo
 \else \expandafter \@secondoftwo
 \fi
}%
\providecommand \natexlab [1]{#1}%
\providecommand \enquote  [1]{``#1''}%
\providecommand \bibnamefont  [1]{#1}%
\providecommand \bibfnamefont [1]{#1}%
\providecommand \citenamefont [1]{#1}%
\providecommand \href@noop [0]{\@secondoftwo}%
\providecommand \href [0]{\begingroup \@sanitize@url \@href}%
\providecommand \@href[1]{\@@startlink{#1}\@@href}%
\providecommand \@@href[1]{\endgroup#1\@@endlink}%
\providecommand \@sanitize@url [0]{\catcode `\\12\catcode `\$12\catcode
  `\&12\catcode `\#12\catcode `\^12\catcode `\_12\catcode `\%12\relax}%
\providecommand \@@startlink[1]{}%
\providecommand \@@endlink[0]{}%
\providecommand \url  [0]{\begingroup\@sanitize@url \@url }%
\providecommand \@url [1]{\endgroup\@href {#1}{\urlprefix }}%
\providecommand \urlprefix  [0]{URL }%
\providecommand \Eprint [0]{\href }%
\providecommand \doibase [0]{https://doi.org/}%
\providecommand \selectlanguage [0]{\@gobble}%
\providecommand \bibinfo  [0]{\@secondoftwo}%
\providecommand \bibfield  [0]{\@secondoftwo}%
\providecommand \translation [1]{[#1]}%
\providecommand \BibitemOpen [0]{}%
\providecommand \bibitemStop [0]{}%
\providecommand \bibitemNoStop [0]{.\EOS\space}%
\providecommand \EOS [0]{\spacefactor3000\relax}%
\providecommand \BibitemShut  [1]{\csname bibitem#1\endcsname}%
\let\auto@bib@innerbib\@empty
\bibitem [{\citenamefont {Day}\ \emph {et~al.}(2003)\citenamefont {Day},
  \citenamefont {LeDuc}, \citenamefont {Mazin}, \citenamefont {Vayonakis},\
  and\ \citenamefont {Zmuidzinas}}]{Day2003}%
  \BibitemOpen
  \bibfield  {author} {\bibinfo {author} {\bibfnamefont {P.~K.}\ \bibnamefont
  {Day}}, \bibinfo {author} {\bibfnamefont {H.~G.}\ \bibnamefont {LeDuc}},
  \bibinfo {author} {\bibfnamefont {B.~A.}\ \bibnamefont {Mazin}}, \bibinfo
  {author} {\bibfnamefont {A.}~\bibnamefont {Vayonakis}},\ and\ \bibinfo
  {author} {\bibfnamefont {J.}~\bibnamefont {Zmuidzinas}},\ }\bibfield  {title}
  {\bibinfo {title} {A broadband superconducting detector suitable for use in
  large arrays},\ }\href {https://doi.org/10.1038/nature02037} {\bibfield
  {journal} {\bibinfo  {journal} {Nature}\ }\textbf {\bibinfo {volume} {425}},\
  \bibinfo {pages} {817} (\bibinfo {year} {2003})}\BibitemShut {NoStop}%
\bibitem [{\citenamefont {Mazin}\ \emph {et~al.}(2013)\citenamefont {Mazin},
  \citenamefont {Meeker}, \citenamefont {Strader}, \citenamefont {Szypryt},
  \citenamefont {Marsden}, \citenamefont {van Eyken}, \citenamefont {Duggan},
  \citenamefont {Walter}, \citenamefont {Ulbricht}, \citenamefont {Johnson},
  \citenamefont {Bumble}, \citenamefont {O'Brien},\ and\ \citenamefont
  {Stoughton}}]{Mazin2013}%
  \BibitemOpen
  \bibfield  {author} {\bibinfo {author} {\bibfnamefont {B.~A.}\ \bibnamefont
  {Mazin}},
  \bibinfo {author} {\bibfnamefont {S.~R.}\ \bibnamefont {Meeker}},
  \bibinfo {author} {\bibfnamefont {M.~J.}\ \bibnamefont {Strader}},
  \bibinfo
  {author} {\bibfnamefont {P.}~\bibnamefont {Szypryt}},
  \bibinfo {author}
  {\bibfnamefont {D.}~\bibnamefont {Marsden}},
  \bibinfo {author} {\bibfnamefont
  {J.~C.}\ \bibnamefont {van Eyken}},
  \bibinfo {author} {\bibfnamefont {G.~E.}\
  \bibnamefont {Duggan}},
  \bibinfo {author} {\bibfnamefont {A.~B.}\
  \bibnamefont {Walter}},
  \bibinfo {author} {\bibfnamefont {G.}~\bibnamefont
  {Ulbricht}},
  \bibinfo {author} {\bibfnamefont {M.}~\bibnamefont {Johnson}},\ \emph{et~al.},\ }\bibfield  {title} {\bibinfo
  {title} {{ARCONS}: A 2024 pixel optical through near-{IR} cryogenic imaging
  spectrophotometer},\ }\href {https://doi.org/10.1086/674013} {\bibfield
  {journal} {\bibinfo  {journal} {Publications of the Astronomical Society of
  the Pacific}\ }\textbf {\bibinfo {volume} {125}},\ \bibinfo {pages} {1348}
  (\bibinfo {year} {2013})}\BibitemShut {NoStop}%
\bibitem [{\citenamefont {Mazin}\ \emph {et~al.}(2009)\citenamefont {Mazin},
  \citenamefont {Young}, \citenamefont {Cabrera},\ and\ \citenamefont
  {Miller}}]{Mazin2009}%
  \BibitemOpen
  \bibfield  {author} {\bibinfo {author} {\bibfnamefont {B.~A.}\ \bibnamefont
  {Mazin}}, \bibinfo {author} {\bibfnamefont {B.}~\bibnamefont {Young}},
  \bibinfo {author} {\bibfnamefont {B.}~\bibnamefont {Cabrera}},\ and\ \bibinfo
  {author} {\bibfnamefont {A.}~\bibnamefont {Miller}},\ }\bibfield  {title}
  {\bibinfo {title} {Microwave kinetic inductance detectors: The first
  decade},\ }\href {https://doi.org/10.1063/1.3292300} {\bibfield  {journal}
  {\bibinfo  {journal} {AIP Conference Proceedings}\ }\textbf {\bibinfo
  {volume} {1185}},\ \bibinfo {pages} {135} (\bibinfo {year}
  {2009})}\BibitemShut {NoStop}%
\bibitem [{\citenamefont {Zobrist}\ \emph {et~al.}(2022)\citenamefont
  {Zobrist}, \citenamefont {Sadlier}, \citenamefont {Daal}, \citenamefont
  {Bumble}, \citenamefont {Day},\ and\ \citenamefont {Mazin}}]{Zobrist2022}%
  \BibitemOpen
  \bibfield  {author} {\bibinfo {author} {\bibfnamefont {N.}~\bibnamefont
  {Zobrist}}, \bibinfo {author} {\bibfnamefont {M.}~\bibnamefont {Sadlier}},
  \bibinfo {author} {\bibfnamefont {M.}~\bibnamefont {Daal}}, \bibinfo {author}
  {\bibfnamefont {B.}~\bibnamefont {Bumble}}, \bibinfo {author} {\bibfnamefont
  {P.~K.}\ \bibnamefont {Day}},\ and\ \bibinfo {author} {\bibfnamefont {B.~A.}\
  \bibnamefont {Mazin}},\ }\bibfield  {title} {\bibinfo {title} {Membraneless
  phonon trapping and resolution enhancement in optical microwave kinetic
  inductance detectors},\ }\href
  {https://doi.org/10.1103/PhysRevLett.129.017701} {\bibfield  {journal}
  {\bibinfo  {journal} {Physical Review Letters}\ }\textbf {\bibinfo {volume}
  {129}},\ \bibinfo {pages} {017701} (\bibinfo {year} {2022})}\BibitemShut
  {NoStop}%
\bibitem [{\citenamefont {Zobrist}\ \emph {et~al.}(2019)\citenamefont
  {Zobrist}, \citenamefont {Daal}, \citenamefont {Bumble}, \citenamefont
  {Baker}, \citenamefont {Subramanian}, \citenamefont {Day},\ and\
  \citenamefont {Mazin}}]{Zobrist2019}%
  \BibitemOpen
  \bibfield  {author} {\bibinfo {author} {\bibfnamefont {N.}~\bibnamefont
  {Zobrist}}, \bibinfo {author} {\bibfnamefont {M.}~\bibnamefont {Daal}},
  \bibinfo {author} {\bibfnamefont {B.}~\bibnamefont {Bumble}}, \bibinfo
  {author} {\bibfnamefont {C.}~\bibnamefont {Baker}}, \bibinfo {author}
  {\bibfnamefont {S.}~\bibnamefont {Subramanian}}, \bibinfo {author}
  {\bibfnamefont {P.}~\bibnamefont {Day}},\ and\ \bibinfo {author}
  {\bibfnamefont {B.~A.}\ \bibnamefont {Mazin}},\ }\bibfield  {title} {\bibinfo
  {title} {Wide-band parametric amplifier readout and resolution of optical
  microwave kinetic inductance detectors},\ }\href
  {https://doi.org/10.1063/1.5098469} {\bibfield  {journal} {\bibinfo
  {journal} {Applied Physics Letters}\ }\textbf {\bibinfo {volume} {115}},\
  \bibinfo {pages} {042601} (\bibinfo {year} {2019})}\BibitemShut {NoStop}%
\bibitem [{\citenamefont {Steiger}\ \emph {et~al.}(2022)\citenamefont
  {Steiger}, \citenamefont {Bailey}, \citenamefont {Zobrist}, \citenamefont
  {Swimmer}, \citenamefont {Dodkins}, \citenamefont {Davis},\ and\
  \citenamefont {Mazin}}]{Steiger2022}%
  \BibitemOpen
  \bibfield  {author} {\bibinfo {author} {\bibfnamefont {S.}~\bibnamefont
  {Steiger}}, \bibinfo {author} {\bibfnamefont {J.~I.}\ \bibnamefont {Bailey},
  \bibfnamefont {III}}, \bibinfo {author} {\bibfnamefont {N.}~\bibnamefont
  {Zobrist}}, \bibinfo {author} {\bibfnamefont {N.}~\bibnamefont {Swimmer}},
  \bibinfo {author} {\bibfnamefont {R.}~\bibnamefont {Dodkins}}, \bibinfo
  {author} {\bibfnamefont {K.~K.}\ \bibnamefont {Davis}},\ and\ \bibinfo
  {author} {\bibfnamefont {B.~A.}\ \bibnamefont {Mazin}},\ }\bibfield  {title}
  {\bibinfo {title} {The {MKID} pipeline: A data reduction and analysis
  pipeline for {UVOIR} {MKID} data},\ }\href
  {https://doi.org/10.3847/1538-3881/ac5833} {\bibfield  {journal} {\bibinfo
  {journal} {The Astronomical Journal}\ }\textbf {\bibinfo {volume} {163}},\
  \bibinfo {pages} {193} (\bibinfo {year} {2022})}\BibitemShut {NoStop}%
\bibitem [{\citenamefont {Zobrist}\ \emph {et~al.}(2021)\citenamefont
  {Zobrist}, \citenamefont {Klimovich}, \citenamefont {Ho~Eom}, \citenamefont
  {Daal}, \citenamefont {Sadlier}, \citenamefont {Bumble}, \citenamefont
  {Day},\ and\ \citenamefont {Mazin}}]{Zobrist2021}%
  \BibitemOpen
  \bibfield  {author} {\bibinfo {author} {\bibfnamefont {N.}~\bibnamefont
  {Zobrist}}, \bibinfo {author} {\bibfnamefont {N.}~\bibnamefont {Klimovich}},
  \bibinfo {author} {\bibfnamefont {B.}~\bibnamefont {Ho~Eom}}, \bibinfo
  {author} {\bibfnamefont {M.}~\bibnamefont {Daal}}, \bibinfo {author}
  {\bibfnamefont {M.}~\bibnamefont {Sadlier}}, \bibinfo {author} {\bibfnamefont
  {B.}~\bibnamefont {Bumble}}, \bibinfo {author} {\bibfnamefont {P.~K.}\
  \bibnamefont {Day}},\ and\ \bibinfo {author} {\bibfnamefont {B.~A.}\
  \bibnamefont {Mazin}},\ }\bibfield  {title} {\bibinfo {title} {Improving the
  dynamic range of single photon counting kinetic inductance detectors},\
  }\href {https://doi.org/10.1117/1.JATIS.7.1.010501} {\bibfield  {journal}
  {\bibinfo  {journal} {Journal of Astronomical Telescopes, Instruments, and
  Systems}\ }\textbf {\bibinfo {volume} {7}},\ \bibinfo {pages} {010501}
  (\bibinfo {year} {2021})}\BibitemShut {NoStop}%
\bibitem [{\citenamefont {Gao}\ \emph {et~al.}(2007)\citenamefont {Gao},
  \citenamefont {Zmuidzinas}, \citenamefont {Mazin}, \citenamefont {LeDuc},\
  and\ \citenamefont {Day}}]{Gao2007}%
  \BibitemOpen
  \bibfield  {author} {\bibinfo {author} {\bibfnamefont {J.}~\bibnamefont
  {Gao}}, \bibinfo {author} {\bibfnamefont {J.}~\bibnamefont {Zmuidzinas}},
  \bibinfo {author} {\bibfnamefont {B.~A.}\ \bibnamefont {Mazin}}, \bibinfo
  {author} {\bibfnamefont {H.~G.}\ \bibnamefont {LeDuc}},\ and\ \bibinfo
  {author} {\bibfnamefont {P.~K.}\ \bibnamefont {Day}},\ }\bibfield  {title}
  {\bibinfo {title} {Noise properties of superconducting coplanar waveguide
  microwave resonators},\ }\href {https://doi.org/10.1063/1.2711770} {\bibfield
   {journal} {\bibinfo  {journal} {Applied Physics Letters}\ }\textbf {\bibinfo
  {volume} {90}},\ \bibinfo {pages} {102507} (\bibinfo {year}
  {2007})}\BibitemShut {NoStop}%
\bibitem [{\citenamefont {Miller}\ \emph {et~al.}(2021)\citenamefont {Miller},
  \citenamefont {Zobrist}, \citenamefont {Ulbricht},\ and\ \citenamefont
  {Mazin}}]{Miller2021}%
  \BibitemOpen
  \bibfield  {author} {\bibinfo {author} {\bibfnamefont {J.~M.}\ \bibnamefont
  {Miller}}, \bibinfo {author} {\bibfnamefont {N.}~\bibnamefont {Zobrist}},
  \bibinfo {author} {\bibfnamefont {G.}~\bibnamefont {Ulbricht}},\ and\
  \bibinfo {author} {\bibfnamefont {B.~A.}\ \bibnamefont {Mazin}},\ }\bibfield
  {title} {\bibinfo {title} {Improving the energy resolution of photon-counting
  microwave kinetic inductance detectors using principal component analysis},\
  }\href {https://doi.org/10.1117/1.JATIS.7.4.048003} {\bibfield  {journal}
  {\bibinfo  {journal} {Journal of Astronomical Telescopes, Instruments, and
  Systems}\ }\textbf {\bibinfo {volume} {7}},\ \bibinfo {pages} {048003}
  (\bibinfo {year} {2021})}\BibitemShut {NoStop}%
\bibitem [{\citenamefont {Fowler}\ \emph {et~al.}(2017)\citenamefont {Fowler},
  \citenamefont {Alpert}, \citenamefont {Doriese}, \citenamefont {Hays-Wehle},
  \citenamefont {Joe}, \citenamefont {Morgan}, \citenamefont {O'Neil},
  \citenamefont {Reintsema}, \citenamefont {Schmidt}, \citenamefont {Ullom},\
  and\ \citenamefont {Swetz}}]{Fowler2016}%
  \BibitemOpen
  \bibfield  {author} {\bibinfo {author} {\bibfnamefont {J.~W.}\ \bibnamefont
  {Fowler}},
  \bibinfo {author} {\bibfnamefont {B.~K.}\ \bibnamefont {Alpert}},
  \bibinfo {author} {\bibfnamefont {W.~B.}\ \bibnamefont {Doriese}},
  \bibinfo
  {author} {\bibfnamefont {J.}~\bibnamefont {Hays-Wehle}},
  \bibinfo {author}
  {\bibfnamefont {Y.~I.}\ \bibnamefont {Joe}},
  \bibinfo {author} {\bibfnamefont
  {K.~M.}\ \bibnamefont {Morgan}},
  \bibinfo {author} {\bibfnamefont {G.~C.}\
  \bibnamefont {O'Neil}},
  \bibinfo {author} {\bibfnamefont {C.~D.}\
  \bibnamefont {Reintsema}},
  \bibinfo {author} {\bibfnamefont {D.~R.}\
  \bibnamefont {Schmidt}},
  \bibinfo {author} {\bibfnamefont {J.~N.}\
  \bibnamefont {Ullom}},\ \emph{et~al.},\ }\bibfield  {title} {\bibinfo {title} {When ``optimal
  filtering'' isn't},\ }\href {https://doi.org/10.1109/TASC.2016.2637359}
  {\bibfield  {journal} {\bibinfo  {journal} {IEEE Transactions on Applied
  Superconductivity}\ }\textbf {\bibinfo {volume} {27}},\ \bibinfo {pages}
  {2100904} (\bibinfo {year} {2017})}\BibitemShut {NoStop}%
\bibitem [{\citenamefont {Gu}\ and\ \citenamefont {Dao}(2024)}]{Gu2023}%
  \BibitemOpen
  \bibfield  {author} {\bibinfo {author} {\bibfnamefont {A.}~\bibnamefont
  {Gu}}\ and\ \bibinfo {author} {\bibfnamefont {T.}~\bibnamefont {Dao}},\
  }\bibfield  {title} {\bibinfo {title} {Mamba: Linear-time sequence modeling
  with selective state spaces},\ }in\ \href@noop {} {\emph {\bibinfo
  {booktitle} {First Conference on Language Modeling (COLM)}}}\ (\bibinfo
  {year} {2024})\ \Eprint {https://arxiv.org/abs/2312.00752} {arXiv:2312.00752
  [cs.LG]} \BibitemShut {NoStop}%
\bibitem [{\citenamefont {Ichinohe}\ \emph {et~al.}(2022)\citenamefont
  {Ichinohe}, \citenamefont {Yamada}, \citenamefont {Hayakawa}, \citenamefont
  {Okada}, \citenamefont {Hashimoto}, \citenamefont {Tatsuno}, \citenamefont
  {Suda},\ and\ \citenamefont {Okumura}}]{Ichinohe2022}%
  \BibitemOpen
  \bibfield  {author} {\bibinfo {author} {\bibfnamefont {Y.}~\bibnamefont
  {Ichinohe}}, \bibinfo {author} {\bibfnamefont {S.}~\bibnamefont {Yamada}},
  \bibinfo {author} {\bibfnamefont {R.}~\bibnamefont {Hayakawa}}, \bibinfo
  {author} {\bibfnamefont {S.}~\bibnamefont {Okada}}, \bibinfo {author}
  {\bibfnamefont {T.}~\bibnamefont {Hashimoto}}, \bibinfo {author}
  {\bibfnamefont {H.}~\bibnamefont {Tatsuno}}, \bibinfo {author} {\bibfnamefont
  {H.}~\bibnamefont {Suda}},\ and\ \bibinfo {author} {\bibfnamefont
  {T.}~\bibnamefont {Okumura}},\ }\bibfield  {title} {\bibinfo {title}
  {Application of deep learning to the evaluation of goodness in the waveform
  processing of transition-edge sensor calorimeters},\ }\href
  {https://doi.org/10.1007/s10909-022-02719-7} {\bibfield  {journal} {\bibinfo
  {journal} {Journal of Low Temperature Physics}\ }\textbf {\bibinfo {volume}
  {209}},\ \bibinfo {pages} {1008} (\bibinfo {year} {2022})}\BibitemShut
  {NoStop}%
\bibitem [{\citenamefont {Fantini}\ and\ \citenamefont {{CUPID
  Collaboration}}(2022)}]{Fantini2022}%
  \BibitemOpen
  \bibfield  {author} {\bibinfo {author} {\bibfnamefont {G.}~\bibnamefont
  {Fantini}}\ and\ \bibinfo {author} {\bibnamefont {{CUPID Collaboration}}},\
  }\bibfield  {title} {\bibinfo {title} {Machine learning techniques for
  pile-up rejection in cryogenic calorimeters},\ }\href
  {https://doi.org/10.1007/s10909-022-02741-9} {\bibfield  {journal} {\bibinfo
  {journal} {Journal of Low Temperature Physics}\ }\textbf {\bibinfo {volume}
  {209}},\ \bibinfo {pages} {1024} (\bibinfo {year} {2022})}\BibitemShut
  {NoStop}%
\bibitem [{\citenamefont {Makhrinov}\ \emph {et~al.}(2025)\citenamefont
  {Makhrinov}, \citenamefont {Maksut}, \citenamefont {Yazici}, \citenamefont
  {Almagambetov}, \citenamefont {Grossan},\ and\ \citenamefont
  {Shafiee}}]{makhrinov2025}%
  \BibitemOpen
  \bibfield  {author} {\bibinfo {author} {\bibfnamefont {V.}~\bibnamefont
  {Makhrinov}}, \bibinfo {author} {\bibfnamefont {Z.}~\bibnamefont {Maksut}},
  \bibinfo {author} {\bibfnamefont {A.}~\bibnamefont {Yazici}}, \bibinfo
  {author} {\bibfnamefont {A.}~\bibnamefont {Almagambetov}}, \bibinfo {author}
  {\bibfnamefont {B.}~\bibnamefont {Grossan}},\ and\ \bibinfo {author}
  {\bibfnamefont {M.}~\bibnamefont {Shafiee}},\ }\bibfield  {title} {\bibinfo
  {title} {Superconducting microresonator signal denoising using machine
  learning},\ }\href {https://doi.org/10.1109/TASC.2024.3513769} {\bibfield
  {journal} {\bibinfo  {journal} {IEEE Transactions on Applied
  Superconductivity}\ }\textbf {\bibinfo {volume} {35}},\ \bibinfo {pages} {1}
  (\bibinfo {year} {2025})}\BibitemShut {NoStop}%
\bibitem [{\citenamefont {Rivasto}\ \emph {et~al.}(2026)\citenamefont
  {Rivasto}, \citenamefont {Isleif}, \citenamefont {Januschek}, \citenamefont
  {Lindner}, \citenamefont {Meyer}, \citenamefont {Othman}, \citenamefont
  {Rubiera~Gimeno},\ and\ \citenamefont {Schwemmbauer}}]{Rivasto2026}%
  \BibitemOpen
  \bibfield  {author} {\bibinfo {author} {\bibfnamefont {E.}~\bibnamefont
  {Rivasto}}, \bibinfo {author} {\bibfnamefont {K.-S.}\ \bibnamefont {Isleif}},
  \bibinfo {author} {\bibfnamefont {F.}~\bibnamefont {Januschek}}, \bibinfo
  {author} {\bibfnamefont {A.}~\bibnamefont {Lindner}}, \bibinfo {author}
  {\bibfnamefont {M.}~\bibnamefont {Meyer}}, \bibinfo {author} {\bibfnamefont
  {G.}~\bibnamefont {Othman}}, \bibinfo {author} {\bibfnamefont {J.~A.}\
  \bibnamefont {Rubiera~Gimeno}},\ and\ \bibinfo {author} {\bibfnamefont
  {C.}~\bibnamefont {Schwemmbauer}},\ }\bibfield  {title} {\bibinfo {title}
  {Binary classification of signal and background triggers of a transition edge
  sensor using convolutional neural networks},\ }\href
  {https://doi.org/10.1038/s41598-025-33353-4} {\bibfield  {journal} {\bibinfo
  {journal} {Scientific Reports}\ }\textbf {\bibinfo {volume} {16}},\ \bibinfo
  {pages} {3389} (\bibinfo {year} {2026})}\BibitemShut {NoStop}%
\bibitem [{\citenamefont {Smith}\ \emph {et~al.}(2024)\citenamefont {Smith},
  \citenamefont {Bailey}, \citenamefont {Cuda}, \citenamefont {Zobrist},\ and\
  \citenamefont {Mazin}}]{Smith2024}%
  \BibitemOpen
  \bibfield  {author} {\bibinfo {author} {\bibfnamefont {J.~P.}\ \bibnamefont
  {Smith}}, \bibinfo {author} {\bibfnamefont {J.~I.}\ \bibnamefont {Bailey},
  \bibfnamefont {III}}, \bibinfo {author} {\bibfnamefont {A.}~\bibnamefont
  {Cuda}}, \bibinfo {author} {\bibfnamefont {N.}~\bibnamefont {Zobrist}},\ and\
  \bibinfo {author} {\bibfnamefont {B.~A.}\ \bibnamefont {Mazin}},\ }\bibfield
  {title} {\bibinfo {title} {{MKIDGen3}: Energy-resolving
  single-photon-counting {MKID} readout on an {RFSoC}},\ }\href
  {https://doi.org/10.1063/5.0225768} {\bibfield  {journal} {\bibinfo
  {journal} {Review of Scientific Instruments}\ }\textbf {\bibinfo {volume}
  {95}},\ \bibinfo {pages} {114705} (\bibinfo {year} {2024})}\BibitemShut
  {NoStop}%
\bibitem [{\citenamefont {Fowler}\ \emph {et~al.}(2015)\citenamefont {Fowler},
  \citenamefont {Alpert}, \citenamefont {Doriese}, \citenamefont {Fischer},
  \citenamefont {Jaye}, \citenamefont {Joe}, \citenamefont {O'Neil},
  \citenamefont {Swetz},\ and\ \citenamefont {Ullom}}]{fowler2015}%
  \BibitemOpen
  \bibfield  {author} {\bibinfo {author} {\bibfnamefont {J.~W.}\ \bibnamefont
  {Fowler}}, \bibinfo {author} {\bibfnamefont {B.~K.}\ \bibnamefont {Alpert}},
  \bibinfo {author} {\bibfnamefont {W.~B.}\ \bibnamefont {Doriese}}, \bibinfo
  {author} {\bibfnamefont {D.~A.}\ \bibnamefont {Fischer}}, \bibinfo {author}
  {\bibfnamefont {C.}~\bibnamefont {Jaye}}, \bibinfo {author} {\bibfnamefont
  {Y.~I.}\ \bibnamefont {Joe}}, \bibinfo {author} {\bibfnamefont {G.~C.}\
  \bibnamefont {O'Neil}}, \bibinfo {author} {\bibfnamefont {D.~S.}\
  \bibnamefont {Swetz}},\ and\ \bibinfo {author} {\bibfnamefont {J.~N.}\
  \bibnamefont {Ullom}},\ }\bibfield  {title} {\bibinfo {title}
  {Microcalorimeter spectroscopy at high pulse rates: A multi-pulse fitting
  technique},\ }\href {https://doi.org/10.1088/0067-0049/219/2/35} {\bibfield
  {journal} {\bibinfo  {journal} {The Astrophysical Journal Supplement Series}\
  }\textbf {\bibinfo {volume} {219}},\ \bibinfo {pages} {35} (\bibinfo {year}
  {2015})}\BibitemShut {NoStop}%
\bibitem [{\citenamefont {Wulf}\ \emph {et~al.}(2020)\citenamefont {Wulf},
  \citenamefont {Jaeckel}, \citenamefont {McCammon}, \citenamefont
  {Chervenak},\ and\ \citenamefont {Eckart}}]{wulf2020}%
  \BibitemOpen
  \bibfield  {author} {\bibinfo {author} {\bibfnamefont {D.}~\bibnamefont
  {Wulf}}, \bibinfo {author} {\bibfnamefont {F.}~\bibnamefont {Jaeckel}},
  \bibinfo {author} {\bibfnamefont {D.}~\bibnamefont {McCammon}}, \bibinfo
  {author} {\bibfnamefont {J.~A.}\ \bibnamefont {Chervenak}},\ and\ \bibinfo
  {author} {\bibfnamefont {M.~E.}\ \bibnamefont {Eckart}},\ }\bibfield  {title}
  {\bibinfo {title} {Optimal filtering of overlapped pulses in microcalorimeter
  data},\ }\href {https://doi.org/10.1063/5.0026193} {\bibfield  {journal}
  {\bibinfo  {journal} {Journal of Applied Physics}\ }\textbf {\bibinfo
  {volume} {128}},\ \bibinfo {pages} {174503} (\bibinfo {year}
  {2020})}\BibitemShut {NoStop}%
\bibitem [{\citenamefont {Gu}\ \emph {et~al.}(2022{\natexlab{a}})\citenamefont
  {Gu}, \citenamefont {Goel},\ and\ \citenamefont {R{\'e}}}]{Gu2022_S4}%
  \BibitemOpen
  \bibfield  {author} {\bibinfo {author} {\bibfnamefont {A.}~\bibnamefont
  {Gu}}, \bibinfo {author} {\bibfnamefont {K.}~\bibnamefont {Goel}},\ and\
  \bibinfo {author} {\bibfnamefont {C.}~\bibnamefont {R{\'e}}},\ }\bibfield
  {title} {\bibinfo {title} {Efficiently modeling long sequences with
  structured state spaces},\ }in\ \href@noop {} {\emph {\bibinfo {booktitle}
  {International Conference on Learning Representations}}}\ (\bibinfo {year}
  {2022})\BibitemShut {NoStop}%
\bibitem [{\citenamefont {Gu}\ \emph {et~al.}(2022{\natexlab{b}})\citenamefont
  {Gu}, \citenamefont {Goel}, \citenamefont {Gupta},\ and\ \citenamefont
  {R{\'e}}}]{Gu2022_S4D}%
  \BibitemOpen
  \bibfield  {author} {\bibinfo {author} {\bibfnamefont {A.}~\bibnamefont
  {Gu}}, \bibinfo {author} {\bibfnamefont {K.}~\bibnamefont {Goel}}, \bibinfo
  {author} {\bibfnamefont {A.}~\bibnamefont {Gupta}},\ and\ \bibinfo {author}
  {\bibfnamefont {C.}~\bibnamefont {R{\'e}}},\ }\bibfield  {title} {\bibinfo
  {title} {On the parameterization and initialization of diagonal state space
  models},\ }in\ \href@noop {} {\emph {\bibinfo {booktitle} {Advances in Neural
  Information Processing Systems}}},\ Vol.~\bibinfo {volume} {35}\ (\bibinfo
  {year} {2022})\BibitemShut {NoStop}%
\bibitem [{\citenamefont {Loshchilov}\ and\ \citenamefont
  {Hutter}(2019)}]{Loshchilov2019}%
  \BibitemOpen
  \bibfield  {author} {\bibinfo {author} {\bibfnamefont {I.}~\bibnamefont
  {Loshchilov}}\ and\ \bibinfo {author} {\bibfnamefont {F.}~\bibnamefont
  {Hutter}},\ }\bibfield  {title} {\bibinfo {title} {Decoupled weight decay
  regularization},\ }in\ \href@noop {} {\emph {\bibinfo {booktitle}
  {International Conference on Learning Representations}}}\ (\bibinfo {year}
  {2019})\ \Eprint {https://arxiv.org/abs/1711.05101} {arXiv:1711.05101
  [cs.LG]} \BibitemShut {NoStop}%
\bibitem [{\citenamefont {Seitzer}\ \emph {et~al.}(2022)\citenamefont
  {Seitzer}, \citenamefont {Tavakoli}, \citenamefont {Antic},\ and\
  \citenamefont {Martius}}]{Seitzer2022}%
  \BibitemOpen
  \bibfield  {author} {\bibinfo {author} {\bibfnamefont {M.}~\bibnamefont
  {Seitzer}}, \bibinfo {author} {\bibfnamefont {A.}~\bibnamefont {Tavakoli}},
  \bibinfo {author} {\bibfnamefont {D.}~\bibnamefont {Antic}},\ and\ \bibinfo
  {author} {\bibfnamefont {G.}~\bibnamefont {Martius}},\ }\bibfield  {title}
  {\bibinfo {title} {On the pitfalls of heteroscedastic uncertainty estimation
  with probabilistic neural networks},\ }in\ \href@noop {} {\emph {\bibinfo
  {booktitle} {International Conference on Learning Representations}}}\
  (\bibinfo {year} {2022})\ \Eprint {https://arxiv.org/abs/2203.09168}
  {arXiv:2203.09168 [cs.LG]} \BibitemShut {NoStop}%
\bibitem [{\citenamefont {Halko}\ \emph {et~al.}(2011)\citenamefont {Halko},
  \citenamefont {Martinsson},\ and\ \citenamefont {Tropp}}]{Halko2011}%
  \BibitemOpen
  \bibfield  {author} {\bibinfo {author} {\bibfnamefont {N.}~\bibnamefont
  {Halko}}, \bibinfo {author} {\bibfnamefont {P.-G.}\ \bibnamefont
  {Martinsson}},\ and\ \bibinfo {author} {\bibfnamefont {J.~A.}\ \bibnamefont
  {Tropp}},\ }\bibfield  {title} {\bibinfo {title} {Finding structure with
  randomness: Probabilistic algorithms for constructing approximate matrix
  decompositions},\ }\href {https://doi.org/10.1137/090771806} {\bibfield
  {journal} {\bibinfo  {journal} {SIAM Review}\ }\textbf {\bibinfo {volume}
  {53}},\ \bibinfo {pages} {217} (\bibinfo {year} {2011})}\BibitemShut
  {NoStop}%
\bibitem [{\citenamefont {Fritsch}\ and\ \citenamefont
  {Carlson}(1980)}]{FritschCarlson1980}%
  \BibitemOpen
  \bibfield  {author} {\bibinfo {author} {\bibfnamefont {F.~N.}\ \bibnamefont
  {Fritsch}}\ and\ \bibinfo {author} {\bibfnamefont {R.~E.}\ \bibnamefont
  {Carlson}},\ }\bibfield  {title} {\bibinfo {title} {Monotone piecewise cubic
  interpolation},\ }\href {https://doi.org/10.1137/0717021} {\bibfield
  {journal} {\bibinfo  {journal} {SIAM Journal on Numerical Analysis}\ }\textbf
  {\bibinfo {volume} {17}},\ \bibinfo {pages} {238} (\bibinfo {year}
  {1980})}\BibitemShut {NoStop}%
\bibitem [{\citenamefont {Kempf}\ \emph {et~al.}(2018)\citenamefont {Kempf},
  \citenamefont {Fleischmann}, \citenamefont {Gastaldo},\ and\ \citenamefont
  {Enss}}]{Kempf2018}%
  \BibitemOpen
  \bibfield  {author} {\bibinfo {author} {\bibfnamefont {S.}~\bibnamefont
  {Kempf}}, \bibinfo {author} {\bibfnamefont {A.}~\bibnamefont {Fleischmann}},
  \bibinfo {author} {\bibfnamefont {L.}~\bibnamefont {Gastaldo}},\ and\
  \bibinfo {author} {\bibfnamefont {C.}~\bibnamefont {Enss}},\ }\bibfield
  {title} {\bibinfo {title} {Physics and applications of metallic magnetic
  calorimeters},\ }\href {https://doi.org/10.1007/s10909-018-1891-6} {\bibfield
   {journal} {\bibinfo  {journal} {Journal of Low Temperature Physics}\
  }\textbf {\bibinfo {volume} {193}},\ \bibinfo {pages} {365} (\bibinfo {year}
  {2018})}\BibitemShut {NoStop}%
\bibitem [{\citenamefont {Ullom}\ and\ \citenamefont
  {Bennett}(2015)}]{Ullom2015}%
  \BibitemOpen
  \bibfield  {author} {\bibinfo {author} {\bibfnamefont {J.~N.}\ \bibnamefont
  {Ullom}}\ and\ \bibinfo {author} {\bibfnamefont {D.~A.}\ \bibnamefont
  {Bennett}},\ }\bibfield  {title} {\bibinfo {title} {Review of superconducting
  transition-edge sensors for x-ray and gamma-ray spectroscopy},\ }\href
  {https://doi.org/10.1088/0953-2048/28/8/084003} {\bibfield  {journal}
  {\bibinfo  {journal} {Superconductor Science and Technology}\ }\textbf
  {\bibinfo {volume} {28}},\ \bibinfo {pages} {084003} (\bibinfo {year}
  {2015})}\BibitemShut {NoStop}%
\bibitem [{\citenamefont {Temples}\ \emph {et~al.}(2024)\citenamefont
  {Temples}, \citenamefont {Wen}, \citenamefont {Ramanathan}, \citenamefont
  {Aralis}, \citenamefont {Chang}, \citenamefont {Golwala}, \citenamefont
  {Hsu}, \citenamefont {Bathurst}, \citenamefont {Baxter}, \citenamefont
  {Bowring}, \citenamefont {Chen}, \citenamefont {Figueroa-Feliciano},
  \citenamefont {Hollister}, \citenamefont {James}, \citenamefont {Kennard},
  \citenamefont {Kurinsky}, \citenamefont {Lewis}, \citenamefont {Lukens},
  \citenamefont {Novati}, \citenamefont {Ren},\ and\ \citenamefont
  {Schmidt}}]{Temples2024}%
  \BibitemOpen
  \bibfield  {author} {\bibinfo {author} {\bibfnamefont {D.~J.}\ \bibnamefont
  {Temples}},
  \bibinfo {author} {\bibfnamefont {O.}~\bibnamefont {Wen}},
  \bibinfo {author} {\bibfnamefont {K.}~\bibnamefont {Ramanathan}},
  \bibinfo
  {author} {\bibfnamefont {T.}~\bibnamefont {Aralis}},
  \bibinfo {author}
  {\bibfnamefont {Y.-Y.}\ \bibnamefont {Chang}},
  \bibinfo {author}
  {\bibfnamefont {S.}~\bibnamefont {Golwala}},
  \bibinfo {author} {\bibfnamefont
  {L.}~\bibnamefont {Hsu}},
  \bibinfo {author} {\bibfnamefont {C.}~\bibnamefont
  {Bathurst}},
  \bibinfo {author} {\bibfnamefont {D.}~\bibnamefont {Baxter}},
  \bibinfo {author} {\bibfnamefont {D.}~\bibnamefont {Bowring}},\ \emph{et~al.},\ }\bibfield  {title} {\bibinfo {title}
  {Performance of a phonon-mediated kinetic inductance detector at the {NEXUS}
  cryogenic facility},\ }\href
  {https://doi.org/10.1103/PhysRevApplied.22.044045} {\bibfield  {journal}
  {\bibinfo  {journal} {Physical Review Applied}\ }\textbf {\bibinfo {volume}
  {22}},\ \bibinfo {pages} {044045} (\bibinfo {year} {2024})}\BibitemShut
  {NoStop}%
\bibitem [{\citenamefont {Mazin}(2026)}]{VENOMcode}%
  \BibitemOpen
  \bibfield  {author} {\bibinfo {author} {\bibfnamefont {B.~A.}\ \bibnamefont
  {Mazin}},\ }\href@noop {} {\bibinfo {title} {{VENOM}: Very efficient neural
  optimal-filter for {MKIDs}}},\ \bibinfo {howpublished} {[Software], GitHub,
  \url{https://github.com/MazinLab/VENOM}} (\bibinfo {year} {2026})\BibitemShut
  {NoStop}%
\bibitem [{\citenamefont {Hannun}\ \emph {et~al.}(2023)\citenamefont {Hannun},
  \citenamefont {Digani}, \citenamefont {Katharopoulos},\ and\ \citenamefont
  {Collobert}}]{MLX2023}%
  \BibitemOpen
  \bibfield  {author} {\bibinfo {author} {\bibfnamefont {A.}~\bibnamefont
  {Hannun}}, \bibinfo {author} {\bibfnamefont {J.}~\bibnamefont {Digani}},
  \bibinfo {author} {\bibfnamefont {A.}~\bibnamefont {Katharopoulos}},\ and\
  \bibinfo {author} {\bibfnamefont {R.}~\bibnamefont {Collobert}},\ }\href@noop
  {} {\bibinfo {title} {{MLX}: Efficient and flexible machine learning on
  {Apple} silicon}},\ \bibinfo {howpublished}
  {\url{https://github.com/ml-explore/mlx}} (\bibinfo {year}
  {2023})\BibitemShut {NoStop}%
\end{thebibliography}%

\end{document}